\documentclass[11pt]{article}
\usepackage{amsbsy,amssymb,latexsym,amsfonts, amsmath,booktabs,tabularx}
\usepackage{mathrsfs}
\usepackage{graphicx}
\usepackage{cite}
\usepackage{comment}

\def\be{\begin{eqnarray}}
\def\ee{\end{eqnarray}}

\newcommand{\dup}{\textup{d}}

\newcommand{\Int}{\int\limits}

\newcommand{\Abs}[1]{\bigl\lvert#1\bigr\rvert}

\newcommand{\xa}[1]{x^{(#1)}}

\catcode`\@=11
\@addtoreset{equation}{section}

\newtheorem{theorem}{Theorem}
\newtheorem{lemma}[theorem]{Lemma}

\newtheorem{prop}{Proposition}

\def\bR{\mathbb{R}}

\newcommand{\nn}{\nonumber}

\def\b{\beta}

\newcommand{\bea}{\begin{eqnarray}}
\newcommand{\eea}{\end{eqnarray}}

\begin{document}
\begin{titlepage}

\begin{center}
\begin{flushright}
UT-11-32
\end{flushright}

\vskip 3 cm
{\huge \bf Selberg Integral and  $SU(N)$ AGT Conjecture}
\vskip 1.3 cm
{\Large Hong Zhang$^\dagger$\footnote{
E-mail address: kilar@hep-th.phys.s.u-tokyo.ac.jp}
and Yutaka Matsuo$^\dagger$\footnote{
E-mail address: matsuo@phys.s.u-tokyo.ac.jp}}
\vskip 1 cm
{\large \sl
$^\dagger$
Department of Physics, The University of Tokyo\\
Hongo 7-3-1, Bunkyo-ku\\
Tokyo 113-0033, Japan}
\end{center}

\vskip 3cm
\begin{abstract}
An intriguing coincidence between the partition function of super
Yang-Mills theory and correlation functions of 2d Toda system
has been heavily studied recently.  While the partition function
of gauge theory was explored by Nekrasov, the correlation functions
of Toda equation have not been completely understood.
In this paper, we study the latter in the form of
Dotsenko-Fateev integral and reduce it in the form of
 Selberg integral of several Jack polynomials.
We conjecture a formula for such Selberg average which satisfies
some consistency conditions and show that it reproduces the $SU(N)$
version of AGT conjecture.
\end{abstract}
\vfill

\end{titlepage}

\tableofcontents
\newpage

\setcounter{footnote}{0}
\section{Introduction}
Two years ago Alday, Gaiotto and Tachikawa \cite{0906.3219} presented an
interesting observation that the partition functions of
certain class of $\mathcal{N}=2$  $SU(2)$ gauge theories
\cite{0306211, 0306238} seem to coincide with
the correlation function of 2D Liouville theory.  After some translation rules
of parameters,
they confirmed a relation which may be written schematically as,
\begin{equation*}
Z^{\mathcal{N}=2}=\left< V\cdots V \right>^{\mathrm{Liouville}}\,.
\end{equation*}
They conjectured that such correspondence exists for large class of
$\mathcal{N}=2$ gauge theories.  Soon later, Wyllard \cite{0907.2189}
and others \cite{0908.2569,Bonelli:2009zp} has presented a
generalization to $SU(N)$ gauge theories.

This conjecture is illuminating in showing a correspondence between
4D Yang-Mills and 2D integrable models and will be fundamental in the understanding
of the duality of gauge theories.   It also will be relevant to understand
strong coupling physics of multiple M5-branes.  In this respect, it will be important
to understand to which extent and how this conjecture holds.
Especially, since the coincidence was found through the first few orders
in the instanton expansion of $q=e^{\pi i \tau}$, the exact computation of conformal
block is needed in the Liouville side.

Recently, A. Mironov et. al. \cite{{1012.3137},{1105.0948}} has embarked on
an interesting step toward this direction.  They used the Dotsenko-Fateev
method \cite{Dotsenko1} to calculate the conformal blocks (see \cite{r:DF,Itoyama}
for earlier contributions).
They analyzed the simplest
example $SU(2)$, $N_f=4$ and proved the AGT relation
for a special choice of a
parameter $\beta=-\epsilon_1/\epsilon_2=1$.  The key step in their analysis
is the reduction of the Dotsenko-Fateev (DF) formula to Selberg average
with one or two Jack polynomial(s) which was computed explicitly
by Kadell \cite{Kadell}.

In this paper, we generalize this idea to $SU(N)$ case.
We find that DF formula is reduced to
$A_{N-1}$-type Selberg average of a product of
$N$ Jack polynomials.
While we do not manage to compute the integral, it is possible to
guess the answer (\ref{nschur}) at least for $\beta=1$.
As we will see, it is still nontrivial task to check if
it reproduces the known results \cite{0708.1193} and satisfies
some consistency conditions that the integral should obey.  With this conjectured
formula, we can prove the $SU(N)$ version of AGT formula.

We organize the sections as follows.  In \S\ref{s:review}, we briefly review
the relevant results of Nekrasov's formula and AGT conjecture.
In \S\ref{s:DF}, we derive the DF formula for the conformal block can reduced
to Selberg integral.  This part is a
generalization of \cite{{1012.3137},{1105.0948}} from $SU(2)$ to $SU(N)$.
In particular, we show how Selberg average of the product of $N$ Jack polynomials
gives the DF formula.  After presenting the known results
\cite{Kadell,0708.1193,0901.4176}, we give a conjecture for  $N$ Jack average
and examine the consistency conditions.
In \S\ref{s:Selberg}, we show that it reproduces the
AGT conjecture properly.

Since this paper needs many technical detail, we have
substantial amount of  sections in the appendix.
In appendix A, we summarize the notation for Young diagrams.
In appendix B, we collect relevant materials on Jack polynomial
which is essential in our computation.  In appendix \ref{a:nJack},
we present the more general conjecture for Selberg integral
for arbitrary $\beta$.  While this formula needs modification, it satisfies
various consistency condition nontrivially and may be useful in the future
development.  In appendix \ref{a:consistency}, we write the explicit computation
of the check of consistency for $N$ Selberg integral.
In appendix E, we give proofs of lemmas which are used to bring Selberg average
into the form of Yang-Mills partition function.

\section{A brief review of AGT conjecture and Nekrasov formula}
\label{s:review}
\paragraph{Nekrasov's partition function}
We first recall the partition function of $\mathcal{N}=2$
super Yang-Mills theory \cite{0306211,0306238}.
With graviphoton deformation parameters $\epsilon_{1},\epsilon_2$
which were introduced for the regularization,
the partition function for
$G=U(N_1)\times \cdots \times U(N_n)$ linear quiver gauge theory
was obtained by the localization technique.
It is schematically written as
\be
 Z_{\mathrm{full}}( q; a, m;\epsilon)= Z_{\mathrm{tree}}
Z_{\mathrm{1loop}}Z_{\mathrm{inst}},\quad
 Z_{\mathrm{inst}}( q; a,  m; \epsilon)=\sum_{\mathbf{Y}}
\mathbf{q}^{\mathbf{Y}} z(\mathbf{Y},  a,  m),
\ee
where
$
\mathbf{Y}:=(\vec Y^{(1)},\cdots, \vec Y^{(n)}),\quad
\mathbf{q}^\mathbf{Y}:=\prod_{i=1}^n q_i^{|\vec Y^{(i)}|}$.
The parameter $ a$ (resp. $m$) represents the diagonalized VEV of vector multiplets
(resp. mass of hypermultiplets) whereas $q_i=e^{\pi i \tau_i}$ is the instanton
expansion parameter for $i$th gauge group $SU(N_i)$.
The total partition function is decomposed into
a product of the contributions of the perturbative parts
$Z_{\mathrm{tree}}$, $Z_{1\mathrm{-loop}}$ and non-perturbative instanton correction
$Z_{\mathrm{inst}}$.
The latter is further decomposed into a sum of sets of
Young diagrams. $\vec Y^{(i)}=(Y_1^{(i)},\cdots, Y_{N_i}^{(i)})$ is a collection of
$N_i$ Young diagram which parameterizes the fixed points of instanton moduli space
for $i$ th gauge group $U(N_i)$.

In this paper, we will mainly focus on the instanton part.
The coefficient $z(\mathbf{Y},  a,  m)$ is described as a
product of the contributions of the gauge- and hyper multiplets
which describes the system:
\be
z(\mathbf{Y},  a, m)=\prod_{i=1}^n
z_{\mathrm{vect}}( a^{(i)}, \vec Y^{(i)})\prod_R z_R(\vec Y, a, m) \; ,
\ee
where $R$ is the representation for each hypermultiplets:
\be
z_{\mathrm{bifund}}(a,\vec Y; b, \vec W;m)&=&
\prod_{t}^{N_1}\prod_{s=1}^{N_2} G_{Y_t, W_s}(a_t-b_s- m)
G_{W_s,Y_t }(b_s - a_t+ m +1 -\beta)\label{bfd} \; ,  \\
z_{\mathrm{fund}}(a,\vec Y; m)&=&\prod_{s=1}^N f_{Y_s}(a_s -m -1 +\beta)\label{fd}\; ,\\
z_{\mathrm{afd}}(a,\vec Y;m)&=&z_\text{fund}( a,\vec Y,-1 +\beta-m)\; ,\\
z_{\mathrm{adj}}(a,\vec Y;m)&=&z_\text{bifund}( a,\vec Y, a,\vec Y,m)\; ,\\
z_{\mathrm{vect}}(a,\vec Y)&=&1/z_\text{adj}( a,\vec Y,0)\; .
\ee
In eq.(\ref{bfd}), the hypermultiplet is supposed to transform as bifundamental
associated with gauge group $U(N_1)\times U(N_2)$.  Similarly, in eq.(\ref{fd}),
the fundamental representation is associated with $U(N)$.
The function $G$ in eq.(\ref{bfd}) is a function with respect to the tableau $Y$'s arm-length and
leg-length(see (\ref{armleg}) for their definitions)
\begin{align}
G_{A,B}(x) = \prod\limits_{(i,j) \in A} \Big(   x + \beta (A'_j - i) + (B_i - j) + \beta \Big) \; ,
\end{align}
and the function $f$ in (\ref{fd}) is defined as
\begin{align}
f_A(z) = \prod\limits_{(i,j) \in A} ( z + \beta ( i - 1) -  (j - 1) ) \; .
\end{align}

Instead of considering general quiver gauge theories,
we are mainly interested in the simplest case, $G=SU(N)$,
with $N_f=2N$ hypermultiplets in fundamental representation.
In this specific example, the partition function is written as
\be
&& Z_{\mathrm{full}}(q; a,  \mu;\epsilon)= Z_{\mathrm{tree}}
Z_{\mathrm{1loop}}Z_{\mathrm{inst}},\quad
 Z_{\mathrm{inst}}( q;a,   m; \epsilon)=\sum_{\vec Y}
q^{|\vec Y|} N^{\mathrm{inst}}_{\vec Y}( a,  \mu),\\
&& N^{\mathrm{inst}}_{\vec Y}(  a,  \mu)=z_{\mathrm{vect}}(\vec Y, a)
\prod_{i=1}^{2N} z_{\mathrm{fund}}(\vec Y, \mu_i)=
 \frac{\prod_{s=1}^{N} \prod_{k=1}^{2N} f_{Y_s}(\mu_k + a_s)}
{\prod_{t,s=1}^{N} g_{Y_t,Y_s}(a_t - a_s)},
\label{Ninst}
\ee
with $g_{AB}(x):=G_{AB}(x) G_{AB}(x+1-\beta)$.
$\mu_i$ ($i=1,\cdots,2N$) are mass parameters of hypermultiplets with
 fundamental representation.

\paragraph{AGT conjecture}
In \cite{0906.3219}
Alday, Gaiotto and Tachikawa pointed out that
this partition function is identical to the
correlation functions of Liouville theory when the gauge group is $SU(2)$.
It takes the form (here we give example of $n$-point function on sphere):
\be
&&\langle V_n(\infty)V_{n-1}(1) V_{n-2}(q_1)\cdots
V_2(q_1\cdots q_{n-3}) V_1(0)\rangle\nonumber\\
&&\quad =\sum_{\psi_1,\cdots,\psi_{n-3}} C_{V_1V_2V_1}
\cdots C_{V_{n-3}V_{n-1}V_{n-1}}
|\mathcal{F}_{V_1V_2U_1\cdots U_{n-3}V_{n-1}V_{n}}(z_1,\cdots,z_n)|^2 \; .
\ee
Here the product of the constants $C_{V_1V_2 U_1}$ etc. are
from the 3-point functions.  For Liouville case, it is given by
DOZZ formula \cite{9403141,9506136,9507109,9911110,0104158,0303150}.
The function $\mathcal{F}$ carries the
coordinate ($q$) dependence and reflects the contributions
of the conformal descendants.  It is called conformal block.

In order to give the identification of partition function with the correlator,
we need some identification of parameters:
$ a,
 m \leftrightarrow \alpha$
and the coordinate $q$ in CFT is identified with
the coupling constant $q=e^{\pi i \tau}$ in Yang-Mills.
Here $\alpha\in \mathbf{R}^N$ is a parameter which appears in the exponential
of the vertex operator  $V_{\alpha}=e^{i(\alpha,\phi)}$
inserted in the correlator.

With such identification, it is shown that $Z_\text{inst}$ in the gauge theory
written in a form \cite{0712.2824} is identical to the conformal blocks,
and the perturbative part $Z_{\mathrm{1loop}}$  corresponds to the
(product of) three point functions \cite{{0906.3219},{0306211},{0306238}}.

To be more explicit, for the specific example of $SU(N)$ gauge theory with
$N_f=2N$ fundamental matter, the relevant Toda
correlator is written in the form 
\be
\langle V_{\alpha_4}(\infty) V_{\alpha_3}(1)
V_{\alpha_2}(q) V_{\alpha_1}(0)\rangle \; ,
\ee
where the insertion of screening operators is necessary for the
charge conservation.
The conformal block of this correlation function is written in the form,
\be
\mathcal{F}_{\alpha_4,\alpha_3,\alpha_2,\alpha_1}(q)=
\sum_{\vec Y} q^{|\vec Y|}
N_{\vec Y}^{\mathrm{Toda}}(\alpha_1,\alpha_2,\alpha_3,\alpha_4)\,.
\ee
It is known that the four point function of Toda theory can
be obtained for special choice of parameters \cite{Fateev:2007ab,Fateev:2008bm},
namely the two of the vertex operator momentum
(say $\alpha_2$ and $\alpha_3$) should be proportional to
either $\omega_1$ or $\omega_{N-1}$ where $\omega_i$ ($i=1,\cdots, N-1$)
is the fundamental weight of $A_{N-1}$.

AGT conjecture for $SU(N)$ \cite{0907.2189,0908.2569} implies that partition function
and the correlator are the same.  In particular it implies,
\be
N^\mathrm{inst}_{\vec Y}(  a, \mu)=N^{\mathrm{Toda}}_{\vec Y}
(\alpha_1,\alpha_2,\alpha_3,\alpha_4),
\ee
if we identify the parameters,
\be\label{e:id}
a= \alpha\ ;\qquad
\mu = -\alpha_1-(1-\beta)\rho,\quad
\tilde \mu = -\alpha_4-(1-\beta)\rho\,;
\ee
where $\mu=(\mu_1,\cdots, \mu_N)$ and $\tilde\mu=(\mu_{N+1},\cdots, \mu_{2N})$
are mass parameters of vector multiplets.
$\alpha=\alpha_1+ \alpha_2 +\beta\sum_a N_a e_a +(1-\beta)=-(\alpha_4+ \alpha_3 +\beta\sum_a \tilde N_a e_a+(1-\beta))$ is the momentum
which appears in the intermediate channel ($N_a$ and $\tilde N_a$ are
the numbers of screening charges and $e_a$ is the simple root
of $A_{N-1}$).  Weyl vector $\rho=\sum_{i=1}^{N-1} \omega_i$
shows up to represent the corrections of the background charge.
As explained, we choose $\alpha_2$ and
$\alpha_3$ to be proportional to $\omega_1$.

We focus on this ``identity" in the following.

\section{Correlation functions of Toda theory and Selberg Formula}
\label{s:DF}

In this section, we give a brief review on Dotsenko-Fateev integral
representation of the correlation function of Toda theory.
We will focus on the four point functions.  We show, by generalizing the
argument of \cite{1012.3137}, that the integral
reduces to the product of Selberg average of $N$ Jack polynomials
for $SU(N)$ Toda theory. Finally, we present our conjecture on
Selberg average which will lead to $SU(N)$ AGT conjecture.

\subsection{$W_N$ algebra and Dotsenko-Fateev integral}
The correlator in $SU(N)$ Toda field theory is given as the
conformal block for $W_{N}$ algebra which consists of the operator algebra
chiral operators $W^{(s)}(z) $ with spin $s= 2,\cdots, N$.
It has a free boson representation
\cite{Fateev:1987zh}.
Let $ \phi(z)=(\phi_1(z),\cdots, \phi_N(z))$ be
free bosons which
satisfies the operator-product expansion\,:
$\phi_j (z) \phi_k(0) \sim \delta^{jk} \log(z)$.
\begin{equation}
R_N= \mathopen{:}\prod_{m=1}^N \left( Q \frac{d}{dz} -
i ( h_m, \partial_z  \phi)\right)\mathclose{:}
= \sum_k  W^{(k)}(z) \left(Q\frac{d}{dz}\right)^{N-k}\,.
\label{miura1}
\end{equation}
$h_m$ are vectors in $\bR^N$ and defined by
$(h_j)_{k}=\delta_{jk}-\frac{1}{N}$.
Since it satisfies $\sum_{m=1}^N (h_j)_m=0$,
a component of $\phi$ is decoupled.
 The definition (\ref{miura1}) gives
 $W^{(0)}(z)=1$ and  $W^{(1)}(z)=0$. The Virasoro generator is
\be
W^{(2)}(z) &=&\frac{1}{2}:(\partial_z\phi)^2: -i Q (\rho, \partial_z^2\phi),
\quad \rho=\sum_{i=1}^{N-1}\omega_i=(\frac{N-1}2,\frac{N-3}2,\cdots,-\frac{N-1}{2}) \; ,
\ee which has the central charge
$c= (N-1) (1 + N(N+1) Q^2)$.

The primary operator of $W_N$ algebra is given as the vertex operators:
\be
V_{\vec\alpha}(z)=:e^{(\alpha, \phi(z))}: \; ,
\ee
which has the OPE with the $W_N$ generators as
\be
W_k(z) V_{\alpha}(0) =\frac{w_k( \alpha)}{z^k} V_{\alpha}(0) +O(z^{-k+1}) \; ,
\ee
with
\be
w_2(\alpha)&=& \Delta(\alpha) =\frac12(\alpha, \alpha)+iQ(\rho,\alpha) \; ,\\
w_k(\alpha)&=&(-1)^k \sum_{1\leq i_1\leq\cdots\leq i_k\leq n}
\prod_{m=1}^k (Q (k-m)+i (h_{i_m}, \alpha))\; .
\ee

In order to derive nonvanishing correlation function of the form
$\langle V_{\vec\alpha_1}(z_1)\cdots V_{\vec\alpha_M}(z_M)\rangle$,
we have freedom to insert screening operators,
\be
Q_j^{(\pm)}=\int \frac{dz}{2\pi i} V_j^{(\pm)} (z) =\int \frac{dz}{2\pi i}
:e^{\alpha_\pm (e_j, \phi(z))}:\,.
\ee
By the requirement of conformal invariance, $w_2(\alpha)=1$,
we need to put $w_2(\alpha_\pm  e_j)=1$.  By writing $Q=i b-i/b$,
the two solutions are $\alpha_+=b$, $\alpha_-=-1/b$.

For the computation of four point functions $\langle V_{\vec \alpha_4} (\infty)
 V_{\vec \alpha_3} (1) V_{\vec \alpha_2} (q) V_{\vec \alpha_4} (0)\rangle$
 we insert $N_a$ screening currents integrated along $[0,q]$ and $\tilde{N}_a$
 currents integrated $[1,\infty]$.
This is a useful prescription to see
the connection with the Selberg formula \cite{Itoyama}.
 For simplicity, we assume we need
 only the screening operators $Q^{(+)}$ in the correlator.
 It gives the Dotsenko-Fateev integral \cite{Dotsenko1} for the four point functions,
\begin{multline}
\label{4point}
Z_\text{DF}(q) = \\
\left\langle \!\!\! \left\langle
: e^{( \tilde{\alpha_1}, \phi(0) )} :: e^{( \tilde{\alpha_2}, \phi(q) )} :
: e^{( \tilde{\alpha_3}, \phi(1) )} :: e^{( \tilde{\alpha_4}, \phi(\infty) )} :
\prod_{a=1}^{N-1} \left(\int_0^q  : e^{b (e_a , \phi (z))}:d z \right)^{N_a}
\left(\int_1^{\infty}  : e^{b (e_a , \phi (z))}:d z \right)^{\tilde{N_a}}
\right\rangle\!\!\!\right\rangle \; .
\end{multline}
For the charge conservation, this correlator has nonvanishing norm
only when
\begin{equation}
\tilde{\alpha_1}+\tilde{\alpha_2}+\tilde{\alpha_3}+\tilde{\alpha_4}+
b\sum_a(N_a+\tilde N_a)e_a
+2iQ \rho=0\,.
\label{cons}
\end{equation}
We apply Wick's theorem to evaluate the correlator
\begin{align}
\left<\left< :e^{(\tilde\alpha_1,\phi(z_1)}): \ \ldots \
:e^{(\tilde\alpha_n,\phi(z_n))}: \right>\right> =
\prod\limits_{ 1 \leq i < j \leq n} (z_j - z_i)^{(\tilde\alpha_i, \tilde\alpha_j)} \; ,
\end{align}
where $e_a$ are the simple roots of $ SU(N)$, and $ (,)$ the bilinear symmetric form on the space dual to
the Cartan subalgebra.
To be consistent with the parameters introduced in the last section,
 defining $\tilde{\alpha_i} = \alpha_i / b $, $\beta =b^2 $ ,
(\ref{4point}) becomes \\
\begin{equation}
\begin{split}
& Z_\text{DF}(q) =q^{(\alpha_1 , \alpha_2)/ \beta} (1-q)^{(\alpha_2 , \alpha_3)/ \beta}
\prod_{a=1}^{N-1} \prod_{I=1}^{N_a}\int_0^q d z_I^{(a)}\prod_{J=N_a +1}^{N_a + \tilde{N_a}}\int_1^{\infty}d z_J^{(a)}
\prod_{i<j}^{N_a + \tilde{N_a}} (z_j^{(a)} - z_i^{(a)})^{2 \beta} \quad \times
\\ &\times
\prod_i^{N_a + \tilde{N_a}} (z_i^{(a)} )^{(\alpha_1 , e_a)}
(z_i^{(a)} - q )^{(\alpha_2 , e_a)} (z_i^{(a)}-1 )^{(\alpha_3 , e_a)}
\prod_{a=1}^{N-2}\prod_i^{N_a + \tilde{N_{a}}} \prod_j^{N_{a+1} + {\tilde N}_{a+1}}(z_j^{(a+1)} - z_i^{(a)})^{- \beta} \quad .
\end{split}\label{ZDF}
\end{equation}
We note that we do not include the 3-point functions in the correlator.
Thus this expression should be compared with the instanton contribution
of Yang-Mills partition functions in AGT conjecture \cite{r:DF}.

\subsection{Reduction to Selberg integral}
In 1944 Selberg find a proof of a noteworthy multiple integral which now plays the role as one of
the most fundamental hypergeometric integrals \cite{1012.3137}.
Here we consider its $A_{N-1}$ extension \cite{0708.1193}
($A_{N-1}$ Selberg integral):
\begin{equation}
S_{\vec u,\vec v,\beta} =\int dx
\prod_{a=1}^{N-1}
\biggl[ \Abs{\Delta\bigl(x^{(a)}\bigr)}^{2\beta}
\prod_{i=1}^{N_a} \bigl(x_i^{(a)}\bigr)^{u_a}
\bigl(1-x_i^{(a)}\bigr)^{v_a}\biggr]
\prod_{a=1}^{N-2}
\Abs{\Delta\bigl(x^{(a)},x^{(a+1)}\bigr)}^{-\beta}
\; ,
\end{equation}
where $\int dx:=\int\limits_{0}^{1} dx^{(1)} \ldots \int\limits_{0}^{1} dx^{(N-1)}$.
As indicated, the integral contains parameters $\vec u=(u_1,\cdots, u_{N-1})$,
$\vec v=(v_1,\cdots, v_{N-1})$ and $\beta$.
Similarly, $A_{N-1}$ Selberg average is the integration with the
Selberg integration kernel,
\begin{multline}
\Big< f \Big>_{\vec u,\vec v,\beta} \ = \ \dfrac{1}{S_{\vec u,\vec v,\beta}}
\int dx
\prod_{a=1}^{N-1}
\biggl[ \Abs{\Delta\bigl(x^{(a)}\bigr)}^{2\beta} \times
\prod_{i=1}^{N_a} \bigl(x_i^{(a)}\bigr)^{u_{a}}
\bigl(1-x_i^{(a)}\bigr)^{v_{a}}\biggr]
\times
\prod_{a=1}^{N-2}
\Abs{\Delta\bigl(x^{(a)},x^{(a+1)}\bigr)}^{-\beta}
  \ f(x) \; .
\end{multline}

In this subsection, we rewrite the Dotsenko-Fateev integral in the form of
$A_{N-1}$ Selberg average for the product of $N$ Jack polynomials
(see appendix \ref{a:Jack} for a summary of relevant material and
\cite{Stanley, Macdonald} for further mathematical details).
In physics literature, Jack polynomial is the eigenfunction of quantum
Calogero-Sutherland model and relevant to the representation theory
of $W_N$ algebra.  See for example \cite{AMOS,r:MY}.
The appearance of the product of $N$ Jack polynomials reminds us of
another line of recent developments \cite{1012.1312,Belavin:2011js,Kanno:2011qv,Fateev:2011hq,Estienne:2011qk}
for the computation of conformal block where the convenient
basis for the Hilbert space is expressed in terms of Jack polynomial.
In particular for $\beta=1$, it is expressed as product of $N$ Schur polynomial.
While the mathematical origin of the appearance of Jack polynomial is different,
there should be a good hint to be learned from each other.

\begin{prop}
The integral (\ref{ZDF})
can be written in the following form (up to $U(1)$ factor),
\begin{equation}
Z_\text{DF}(q)=
\sum_{\vec{Y}}
 q^{\lvert \vec{Y} \rvert}
\left\langle
\prod_{a=1}^{N}
j^{(\beta)}_{Y_a} (-r_k^{(a)} - \frac{v'_{a+}}{\beta})
\right\rangle_{+}
\left\langle
\prod_{a=1}^{N}
j^{(\beta)}_{Y_a} (\tilde{r}_k^{(a)}+ \frac{v'_{a-}}{\beta})
\right\rangle_{-} \; .
\label{DF-S}
\end{equation}
Here we have to explain some notations.
$\vec Y$ is a collection of $N$ Young diagrams,
$j^{(\beta)}_Y$ is normalized Jack symmetric polynomial.
We introduced new parameters $v_{a\pm}$ and $u_{a\pm}$ by
\begin{equation}
v_{a+}=(\alpha_2 , e_a), \quad v_{a-}=(\alpha_3 , e_a),\quad u_{a+}=(\alpha_1 , e_a),
\quad u_{a-}=(\alpha_4, e_a),
\end{equation}
where we use a relation
\begin{equation}
u_{a+}+u_{a-}+v_{a+}+v_{a-}+\beta\sum_b C_{ab}(N_b+\tilde N_b) =2\beta -2
\end{equation}
implied by Eq.(\ref{cons}) to define $u_{a-}$.
The Selberg average $\langle\cdots\rangle_\pm$ is taken
with respect to these parameters,
$\langle\cdots\rangle_{\pm}:=\langle\cdots\rangle_{\vec u_\pm,\vec v_\pm,\beta}$.
$r_k^{(a)}$ and $\tilde r_k^{(a)}$ is related to the integration variables $x^{(a)}_i$
and $y^{(a)}_i$ through
\begin{eqnarray}
r_k^{(a)}:=p_k^{(a)}-p_k^{(a-1)},\quad p_k^{(a)}:=\sum_i(x^{(a)}_i)^k\quad
\mbox{and}\quad
 \tilde r_k^{(a)}:=\tilde p_k^{(a)}-\tilde p_k^{(a-1)},\quad
\tilde p_k:=\sum_i(y^{(a)}_i)^k\,,
\end{eqnarray}
with $p_k^{(0)}=p_k^{(N)}=\tilde p_k^{(0)}=\tilde p_k^{(N)}=0$.
Finally
$v'_{a-}:=-\sum_{s=1}^{a-1} v_{s-}$, and $ v'_{(N-a)+} :=  \sum_{s=1}^{a}v_{(N-s)+}$.
\end{prop}

In particular, when $N=2$, the above reduce to (notice that $ v'_{1-}= v'_{2+} = 0$)
\begin{equation}
Z_{DF}(q)=\sum_{A,B} q^{\lvert A \rvert + \lvert B \rvert}
\left\langle
j^{(\beta)}_{A} (-p_k-\frac{v_+}{\beta}) j^{(\beta)}_{B} (p_k)
\right\rangle_{+}
\left\langle
 j^{(\beta)}_{A} (\tilde {p}_k)j^{(\beta)}_{B} (-\tilde {p}_k-\frac{v_-}{\beta})
\right\rangle_{-}\;,
\end{equation}
\\
which was used in \cite{1012.3137}.  The proposition is
a generalization of their result.

\paragraph{Proof:}
Let us derive the proposition in the rest of this subsection.
Following the procedure in \cite{Itoyama} for $SU(2)$,
we rename the integration variables in (\ref{ZDF})
$z_I =: q x_I$, $1\leq I \leq N_a$ and $z_J =: \frac{1}{y_J} $, $N_a + 1\leq J \leq N_a + \tilde{N_a} $.
Then Eq.\eqref{ZDF} is rewritten as a double average\footnote{
The $U(1)$ prefactors are omitted for its
irrelevance to the Nekrasov function.},
\begin{equation}\label{double}
\left\langle \!\!\! \left\langle
\prod_{a=1}^{N-1}
\Bigg\{
\prod_{i=1}^{N_a} (1-q x_i^{(a)})^{v_{a-}}
\prod_{j=1}^{\tilde{N_a}} (1-q y_j^{(a)})^{v_{a+}}
\Bigg\}
\prod_{a=1}^{N-1}\prod_{b=1}^{N-1}
\Bigg\{
\prod_{i=1}^{N_a}\prod_{j=1}^{\tilde{N_b}}(1-q x_i^{(a)} y_j^{(b)})^{C_{ab} \beta}
\Bigg\}
\right\rangle_{+} \right\rangle_{-} \; ,
\end{equation}
where $C_{ab}$ is $A_{N-1}$ Cartan matrix,
\[ C_{ab}=\begin{cases}
2 & a=b\\
-1 & a=b \pm 1 \\
0 & |a-b| > 1 \; ,
\end{cases} \]
and
the Selberg average $\langle \cdots\rangle_+$ (resp. $\langle \cdots\rangle_-$)
is taken over the variables $x_i^{(a)}$ (resp. $y_i^{(a)}$) with parameters
$\vec u_+, \vec v_+$ (resp. $\vec u_-, \vec v_-$).

We change the second product in the integral \eqref{double}
into exponential form
\be
\prod_{a,b=1}^{N-1}
\prod_{i=1}^{N_a}\prod_{j=1}^{\tilde{N_b}}(1-q x_i^{(a)} y_j^{(b)})^{C_{ab} \beta}
&=&
\text{exp}\Bigg\{ \beta \sum_{a,b=1}^{N-1} C_{ab} \sum_{i,j}\ln(1-q x_i^{(a)} y_j^{(b)} )\Bigg\} \nn\\
&=&
\text{exp}\Bigg\{ - \beta \sum_{a,b=1}^{N-1} C_{ab} \sum_{k=1}^{\infty}\frac{q^k}{k}p_k^{(a)} \tilde{p}_k^{(b)} \Bigg\}\nn
\\
&=&
\text{exp} \Bigg\{ - \beta  \sum_{k=1}^{\infty}\frac{q^k}{k}
\Bigg[ 2 \sum_{a=1}^{N-1} p_k^{(a)} \tilde{p}_k^{(a)}
-\sum_{a=2}^{N-1} p_k^{(a)} \tilde{p}_k^{(a-1)}
-\sum_{a=1}^{N-2} p_k^{(a)} \tilde{p}_k^{(a+1)}
\Bigg ] \Bigg\}\nn\\
&=&
\text{exp} \Bigg\{ - \beta  \sum_{k=1}^{\infty}\frac{q^k}{k}
\sum_{a=1}^{N} r_k^{(a)} \tilde{r}_k^{(a)}
\Bigg\} \; .
\ee
In the second line, we performed Taylor expansion
and rewrite the variables $x,y$  by $p_k^{(a)}$ and $\tilde{p}_k^{(b)}$.
In the last line, we rewrite $p_k, \tilde p_k$ by
$ r_k^{(a)}, \tilde r_k^{(a)} $.

Likewise, we rewrite
\be
\prod_{a=1}^{N-1}
\prod_{i=1}^{N_a} (1-q x_i^{(a)})^{v_{a-}}
=
\text{exp} \Bigg\{ - \beta  \sum_{k=1}^{\infty}\frac{q^k}{k}
\sum_{a=1}^{N-1} p_k^{(a)} \frac{v_{a-}}{\beta}
\Bigg\}
\equiv
\text{exp} \Bigg\{ - \beta  \sum_{k=1}^{\infty}\frac{q^k}{k}
\sum_{a=1}^{N} r_k^{(a)} \frac{v'_{a-}}{\beta}
\Bigg\} \; .
\ee
In the second equivalence we change the basis from $ p_k^{(a)}$ to $ r_k^{(a)}$.
The coefficients $v'_{a-}$ are determined from $ v_{a-}$ with an additional
condition $ v'_{1-}:=0$ which is somewhat arbitrary.
Similarly,
\begin{equation}
\prod_{a=1}^{N-1}
\prod_{j=1}^{\tilde{N_a}} (1-q y_j^{(a)})^{v_{a+}}
=
\text{exp}\Bigg\{ - \beta  \sum_{k=1}^{\infty}\frac{q^k}{k}
\sum_{a=1}^{N} \tilde{r}_k^{(a)} \frac{v'_{a+}}{\beta}
\Bigg\} \; .
\end{equation}
This time we define $v'_{a+}$ from another condition
$ v'_{N+}=0$ for the convenience of later arguments. \\ \\
Combining the above factors together, the integrand in \eqref{double}
takes the form
\begin{multline}
\text{exp} \Bigg\{ - \beta  \sum_{k=1}^{\infty}\frac{q^k}{k}
\sum_{a=1}^{N}\Big[ (r_k^{(a)} + \frac{v'_{a+}}{\beta})(\tilde{r}_k^{(a)}+ \frac{v'_{a-}}{\beta})
-\frac{v'_{a+}}{\beta} \frac{v'_{a-}}{\beta}
\Big]
\Bigg\}
\\
=
\prod_{a=1}^{N} (1-q)^{v'_{a+} v'_{a-} / \beta}
\sum_{\vec{Y}}\prod_{a=1}^{N}
 q^{\lvert \vec{Y} \rvert}
j_{Y_a} (-r_k^{(a)} - \frac{v'_{a+}}{\beta}) j_{Y_a} (\tilde{r}_k^{(a)}+ \frac{v'_{a-}}{\beta}) \;,
\end{multline}
where we have made use of the Cauchy-Stanley identity (\ref{eCR})
for the Jack polynomial in the second line
\begin{equation}
\text{exp}(\beta  \sum_{k=1}^{\infty}\frac{1}{k}p_k p'_k)=\sum_R j^{(\beta)}_{R}(p)j^{(\beta)}_{R}(p')\,.
\end{equation}
So the conformal blocks (\ref{4point}) finally becomes
\begin{equation}
\prod_{a=1}^{N} (1-q)^{v'_{a+} v'_{a-} / \beta}
\sum_{\vec{Y}}
 q^{\lvert \vec{Y} \rvert}
\left\langle
\prod_{a=1}^{N}
j^{(\beta)}_{Y_a} (-r_k^{(a)} - \frac{v'_{a+}}{\beta})
\right\rangle_{+}
\left\langle
\prod_{a=1}^{N}
j^{(\beta)}_{Y_a} (\tilde{r}_k^{(a)}+ \frac{v'_{a-}}{\beta})
\right\rangle_{-} \;.
\end{equation}
Absorbing the prefactor into the $U(1)$ part of the product, we arrive at
(\ref{DF-S}).
{\bf QED}

\subsection{Known results and a conjecture on Selberg average}
\label{s:conj}
The Dotzenko-Fateev integral is now reduced to the evaluation of
Selberg average of $N$ Jack polynomials. Let us first summarize the known results
on Selberg average in the literature.
\paragraph{$SU(2)$ case:}
The relevant
Selberg averages for one and two Jack polynomials were obtained by Kadell \cite{Kadell},
\begin{eqnarray}
&& \Big< J_Y^{(\beta)}(p) \Big>^{{SU(2)}}_{u,v,\beta} \
= \dfrac{[N\beta]_{Y}[u + N\beta + 1 - \beta]_Y}
{\prod\limits_{(i,j) \in Y} \Big( \beta ( Y^{\prime}_j - i) + (Y_i - j)
+ \beta \Big)[u+v+2N\beta + 2 - 2\beta]_Y}\;,\\
&&\Big< J_A^{(\beta)}(p + w) J_B^{(\beta)}(p) \Big>^{{SU(2)}}
= \  \dfrac{[v+N\beta+1-\beta]_A [u+N\beta+1-\beta]_B}{[N \beta]_A
[u+v+N\beta+2-2\beta]_B} \times\label{2jack}
 \\
&& ~~~~ \times
\dfrac{\prod\limits_{i<j}^{N} \Big( A_i - A_j + (j-i)\beta\Big)_{\beta} \prod\limits_{i<j}^{N} \Big( B_i - B_j + (j-i)\beta\Big)_{\beta} }
{\prod\limits_{i,j}^{N} \Big( \ u+v+2\beta N + 2+A_i+B_j - (1+i+j)\beta \ \Big)_{\beta}}
\times
\dfrac{\prod\limits_{i,j}^{N} \Big( \ u+v+2\beta N + 2 - (1+i+j)\beta \ \Big)_{\beta}}
{\prod\limits_{i<j}^{N} \Big( (j-i)\beta\Big)_{\beta} \prod\limits_{i<j}^{N} \Big(  (j-i)\beta\Big)_{\beta} }\nonumber \;,
\end{eqnarray}
where we have used the following notation
\begin{equation}
\label{trick0}
[ x ]_A
=
\prod\limits_{(i,j) \in A} ( x - \beta ( i - 1) +  j - 1 )
=(-1)^{|A|}f_A(-x)\,,
\end{equation}
and Pochhammer symbol
\begin{align}
\label{poch}
(x)_{k} = \dfrac{\Gamma(x+k)}{\Gamma(x)} = x(x+1)\ldots(x+k-1)\,.
\end{align}
$J_Y^{(\beta)}$, the Jack polynomial, is related to normalized one
$j_Y^{(\beta)}$ as (\ref{jj}).
Inclusion of a shift $w$ of the argument for the two Jack case was
conjectured in \cite{1012.3137}.
Together with the identity
$
j^{(\beta)}_{A}(-p/\beta) = (-1)^{|A|} j^{(1/\beta)}_{A^\prime}(p)
$ and an identification of parameter $w = (v+1 -\beta)/ \beta$,
these are sufficient to evaluate (\ref{DF-S}) for $SU(2)$ case \cite{1012.3137}.

\paragraph{$SU(n+1)$ case:}
The one-Jack Selberg integral for $SU(n+1)$ could
be calculated by the formula offered by
Warnaar \cite{0708.1193}.  To perform the integral,
we need to restrict the parameter $v$ as,
\begin{equation}\label{e:constr}
v_2=\cdots=v_{n}=0,\quad \mbox{and}\quad v_1=v.
\end{equation}
As already explained,
in Toda field theory, this condition is necessary to solve conformal Ward
identity for the W-algebra \cite{Fateev:2007ab, 0907.2189}.
The formula by Warnaar is,
\begin{eqnarray}
\label{n1jacks}
&& \left\langle J_{B}^{(\beta)} (p_k^{(n)})  \right\rangle^{SU(n+1)}_{\vec u,\vec v,\beta}
=\prod_{1\leq i<j\leq N_n}\frac{((j-i+1)\beta)_{B_i-B_j}}
{((j-i)\beta)_{B_i-B_j}} \times \nonumber\\
&&\quad\times
\prod_{a=1}^n
\prod_{i=1}^{N_n}
\frac{(u_{n-a+1}+\cdots+u_n+a+(N_n-a-i+1)\beta)_{B_i}}
{(v_{n-a+1}+u_{n-a+1}+\cdots+u_n+a+1+(N_n+N_{n-a+1}-N_{n-a}-a-i)\beta)_{B_i}} \;.
\end{eqnarray}
To evaluate (\ref{DF-S}), we need Selberg average of $(n+1)$ Jack polynomials.
While we do not perform the integration so far, we find a formula for
$\beta=1$ which reproduces known results and satisfies consistency conditions\footnote{
Actually we could guess a formula for general $\beta$
(see appendix \ref{a:nJack}) which reproduces the known results.
While the formula looks quite reasonable, it does not pass one of the consistency checks.
It seems that some modifications up to the terms proportional to $1-\b$ are needed.}.
As explained in appendix \ref{a:Jack}, the Jack polynomial for $\beta=1$
is called Schur polynomial and we write $J^{(\beta)}_Y|_{\beta=1}=\chi_Y$.

\paragraph{Conjecture }
{\em We propose the following formula of Selberg average for $n+1$ Schur
polynomials,
} 
{\small
\begin{equation}
\begin{split}
\label{nschur}
&\left\langle
\chi_{Y_1} (-p_k^{(1)} - v_1')\dots
\chi_{Y_r} (p_k^{(r-1)} -p_k^{(r)} - v_r')\dots
\chi_{Y_{n+1}} (p_k^{(n)})
 \right\rangle^{SU(n+1)}_{\vec u,\vec v,\beta=1}
\\
&=\prod_{s=1}^{n}
\bigg\{
(-1)^{|Y_s|}\; \times
\frac{[v_{s}+ N_s - N_{s-1} ]_{Y'_s} }
{[ N_s + N_{s-1}  ]_{Y'_s} }
 \; \times \!
\prod_{1\leq i<j\leq N_{s-1} + N_s}
\frac{(j-i+1)_{Y'_{si}-Y'_{sj}}}{(j-i)_{Y'_{si}-Y'_{sj}}}
\bigg\}
\quad\times
\prod_{1\leq i<j\leq N_{n}}
\frac{(j-i+1)_{Y_{(n+1)i}-Y_{(n+1)j}}}{(j-i)_{Y_{(n+1)i}-Y_{(n+1)j}}}
\\
&\times
\prod_{1\leq t<s\leq n+1}
\bigg\{
\frac{[ v_{t}+u_{t}+\dots +u_{s-1} + N_t- N_{t-1}   ]_{Y'_t} }
{[ v_{t}- v_{s}+u_{t}+\dots +u_{s-1}+ N_{t} - N_{t-1} - N_s  ]_{Y'_t} }
\times
\frac{[-v_{s}+u_{t}+\dots +u_{s-1} - N_s + N_{s-1} ]_{Y_s} }
{[ v_{t}- v_{s}+u_{t}+\dots +u_{s-1} - N_{t-1} - N_s + N_{s-1}  ]_{Y_s} }
 \\
&\qquad\times
\prod_{i=1}^{N_{t}}\prod_{j=1}^{N_{s-1}}
\frac{  v_{t}- v_{s}+u_{t}+\dots +u_{s-1} + N_t- N_{t-1} - N_s + N_{s-1}  +1 -(i +j)}
{ v_{t}- v_{s}+u_{t}+\dots +u_{s-1} + N_t- N_{t-1} - N_s + N_{s-1}   +1 + Y'_{ti} + Y_{sj} -(i +j) }
\bigg\} \;,
\end{split}
\end{equation}
with $v_r':=\sum_{a=r}^n v_a=v\delta_{r1}$ after imposing the constraint (\ref{e:constr}).
}

As we wrote, this formula seems reasonable since
\begin{itemize}
\item It reproduces the AGT relation as we will see in the next section.
\item It is reduced to the known results for $\beta=1$
with the help of (\ref{trick5}),
\begin{itemize}
\item[(a)] For $Y_{1} = \dots = Y_{n} = \emptyset$, and $Y_{n+1} = B$,
 the above reduce to the $A_n$
one Jack integral (\ref{n1jacks}).  
\item[(b)] For $n= 1$, $Y_{1} = A$ and $Y_{2} = B$,
 it coincides with the $A_1$ two Jack integral (\ref{2jack}).
\item[(c)] For $n= 2$, $Y_{1} = R$, $Y_{2} = \emptyset$, and $Y_{3} = B$,
the above is consistent with the $A_2$
two Jack integral (\ref{2jacka2}) given by Warnaar \cite{0901.4176}.
\item[(d)] For $N_n =0$, $u_n=0$ and $Y_{n+1} = \emptyset$, the above reduces to
the formula for $A_{n-1}$.
\end{itemize}
\end{itemize}

Another type of consistency conditions is also considered.
For the simplest case,
we start from multiplying a trivial zero factor
$$
v+(-p^{(1)}_1-v)+(p^{(1)}_1-p^{(2)}_1)+\cdots+(p^{(n-1)}_1-p^{(n)}_1)+p^{(n)}_1=0
$$
in the integrand of (\ref{nschur}).
We then apply to each term a property of Schur polynomial,
\be\label{schuid}
p_1 \chi_R(p_k)=\sum_{\tilde{R}} \chi_{\tilde{R}}(p_k)\;,
\ee
where the summation is over all possible Young diagrams which can be obtained
from $R$ by adding one cell.  This gives rise to a consistency condition
for any combination $(Y_1,\cdots,Y_{n+1})$;
\be
\label{conscheck}
&& v\left\langle
\chi_{Y_1} (-p_k^{(1)} - v_1')\dots
\chi_{Y_r} (p_k^{(r-1)} -p_k^{(r)} - v_r')\dots
\chi_{Y_{n+1}} (p_k^{(n)})
 \right\rangle^{SU(n+1)}_{\vec u,\vec v,\beta=1}\nonumber\\
 && ~~~~~+\sum_{r=1}^{n+1} \sum_{\tilde{Y}_r}
 \left\langle
\chi_{Y_1} (-p_k^{(1)} - v_1')\dots
\chi_{\tilde{Y}_r} (p_k^{(r-1)} -p_k^{(r)} - v_r')\dots
\chi_{Y_{n+1}} (p_k^{(n)})
 \right\rangle^{SU(n+1)}_{\vec u,\vec v,\beta=1}=0\;.
\ee
While this looks trivial, the cancellation becomes rather nontrivial.
We give a detailed computation for the simpler cases,
$n=2$ ($SU(3)$) with $Y_1, Y_2, Y_3$ being rectangle Young diagrams,
in appendix \ref{a:consistency}.

We may write easily some generalizations of (\ref{schuid}) such as,
\be
\chi_{[n]}(p_k) \chi_R(p_k)=\sum_{\tilde{R}} \chi_{\tilde{R}}(p_k) \;,
\ee
where $\tilde{R}/R$ is $[n]$.  We hope that such series of consistency conditions may
serve as a proof of the formula (\ref{nschur}) in the future.

\section{AGT conjecture from Selberg integral}
\label{s:Selberg}

In the following, we present a `proof' of AGT conjecture
for $SU(n+1)$ case by using the postulated formulae for Selberg average
in \S\ref{s:conj}.  It is a generalization of the proof for
 $SU(2)$ case in \cite{{1012.3137},{1105.0948}}.
As we already mentioned, what we need to see is the coincidence of
partition function,
\be
Z_\text{inst}(q)=Z_\text{DF}(q) \;,
\ee
up to $U(1)$ factor but we would like to see the stronger condition,
namely the coefficient $N^\mathrm{Inst}$ in the instanton partition function
(\ref{Ninst}) with the similar coefficient  $N^\mathrm{Toda}$ in
(\ref{DF-S})
\be\label{AGT1}
N_{\vec Y}^\mathrm{inst}=N_{\vec Y}^\mathrm{Toda}\,.
\ee
We show that this stronger identity
holds at $\beta=1$.

We note that both coefficients have the factorized form:
\begin{equation}
N_{\vec Y}^\mathrm{inst}\equiv
N_{\vec Y +}^\mathrm{inst}
N_{\vec Y -}^\mathrm{inst},\quad
N_{\vec Y}^\mathrm{Toda}\equiv
N_{\vec Y +}^\mathrm{Toda}
N_{\vec Y -}^\mathrm{Toda},
\end{equation}
with
\be
N_{\vec Y +}^\mathrm{inst}&\equiv&
\frac{\prod_{s=1}^{n+1}
\prod_{k=1}^{n+1} f_{Y_s}(\mu_k + a_s)}
{\prod_{t,s=1}^{n+1} G_{Y_t,Y_s}(a_t - a_s) }
\prod_{s=1}^{n+1}
\bigg\{
(-1)^{|Y_s|}
\sqrt{\frac{G_{Y_s,Y_s}(0)}{G_{Y_s,Y_s}(1-\beta)}}
\bigg\}
\,,\nonumber\\
N_{\vec Y -}^\mathrm{inst}&\equiv&
 \frac{\prod_{s=1}^{n+1} \prod_{k=n+2}^{2n+2} f_{Y_s}(\mu_k + a_s)}
{\prod_{t,s=1}^{n+1}G_{Y_t,Y_s}(a_t - a_s +1 - \beta)}
\prod_{s=1}^{n+1}
\bigg\{
(-1)^{|Y_s|}
\sqrt{\frac{G_{Y_s,Y_s}(1-\beta)}{G_{Y_s,Y_s}(0)}}
\bigg\} \;,
\label{ninst}
\ee
and
\begin{equation}
N_{\vec Y \pm}^\mathrm{Toda}\equiv
\left\langle
\prod_{a=1}^{n+1}
j^{(\beta)}_{Y_a} (-r_k^{(a)} - \frac{v'_{a\pm}}{\beta})
\right\rangle_{\pm}
=
\prod_{a=1}^{n+1}\sqrt{\frac{G_{Y_a,Y_a}(0)}{G_{Y_a,Y_a}(1-\beta)}}
\left\langle
\prod_{a=1}^{n+1}
J^{(\beta)}_{Y_a} (-r_k^{(a)} - \frac{v'_{a\pm}}{\beta})
\right\rangle_{\pm}\,.
\end{equation}
We remind that $ r_k^{(a)} \equiv p_k^{(a)} - p_k^{(a-1)} $ ,
$ v'_{a-} = - \sum_{s=1}^{a-1}v_{s-}$ and
$ v'_{(N-a)+} =\sum_{s=1}^{a}v_{(N-s)+}$.
Therefore, the problem left is to figure out whether the
(n+1)-Jack Selberg integral has
 the same form with its Nekrasov counterpart
 for $\beta=1$,
\begin{equation}
N_{\vec Y \pm}^\mathrm{Toda}
=
N_{\vec Y \pm}^\mathrm{inst}\quad.
\label{AGTb}
\end{equation}

\subsection{Special case: $\vec Y=(\emptyset,\cdots,\emptyset,B)$, arbitrary
$\beta$}
In the following, we prove (\ref{AGTb}) for `$+$' part.
Proof for `$-$' is similar.
We will omit the lower index"+" in $v_{a+}$and $u_{a+}$
as long as there are no misunderstanding.

We start from the simplest case, when
$Y_1=\dots =Y_n= \emptyset ,\; Y_{n+1}=B$.
In this case, the Selberg integral is already proved by Warnaar
for arbitrary $\beta$.
So our proof for this case is exact and holds without the restriction
of $\beta$.

In the instanton part, we have,
\begin{equation}
\label{n1jacki}
N_{(\emptyset ,\dots, \emptyset , B) +}^\mathrm{inst}
= \frac{(-1)^{|B|} \prod_{k=1}^{n+1} f_{B}(\mu_k + a_{n+1})}
{\sqrt{G_{B,B}(0)G_{B,B}(1-\beta)} \prod_{m=1}^{n}G_{B,\emptyset}(a_{n+1} - a_m)} \;.
\end{equation}
On the other hand, the one-Jack Selberg integral is given in (\ref{n1jacks})
\begin{equation}\label{n1t}
\begin{split}
N_{(\emptyset ,\dots, \emptyset , B) +}^\mathrm{Toda}&=
\left\langle
j^{(\beta)}_{B} (p_k^{(n)})  \right\rangle_{+}^{SU(n+1)}
 \\
&=\sqrt{\frac{G_{B,B}(0)}{G_{B,B}(1-\beta)}} \times
\left\langle
J_{B} (p_k^{(n)})  \right\rangle_{+}^{SU(n+1)}
 \\
&=\sqrt{\frac{G_{B,B}(0)}{G_{B,B}(1-\beta)}} \times
\prod_{1\leq i<j\leq N_n}\frac{((j-i+1)\beta)_{B_i-B_j}}
{((j-i)\beta)_{B_i-B_j}} \\
&\quad\times
\prod_{a=1}^n
\prod_{i=1}^{N_n}
\frac{(u_{n-a+1}+\cdots+u_n+a+(N_n-a-i+1)\beta)_{B_i}}
{(v_{n-a+1}+u_{n-a+1}+\cdots+u_n+a+1+(N_n+N_{n-a+1}-N_{n-a}-a-i)\beta)_{B_i}} \;.
\end{split}
\end{equation}

To see the equivalence, first we note that the function $f_B(x)$ in $N^\mathrm{inst}$ is linked to the notation
$[x]_B$ by (\ref{trick0}). Then we need to rewrite $G_{AB}$ in terms of $(x)_B$
in (\ref{n1t}). For this purpose, we need the following lemmas which will be
proved in appendix:
\begin{lemma}
\begin{equation}
\label{trick1}
\prod_{1\leq i<j\leq N}\frac{((j-i+1)\beta)_{B_i-B_j}}
{((j-i)\beta)_{B_i-B_j}}
=
\frac{[ N \beta ]_B}{G_{B,B}(0) }
\end{equation}
\end{lemma}

\begin{lemma}
\begin{equation}
\label{trick2}
\prod_{i=1}^{N} (x - i \beta)_{B_i}
=
\Big[ x - \beta \Big]_B
\end{equation}
\end{lemma}
\begin{lemma}
\begin{equation}
\label{trick3}
[ x ]_B
=
(-1)^{|B|} G_{B,\emptyset}(- x + 1 -\beta)
\end{equation}
\end{lemma}
With the help of these formulae,
we arrive at the results
\begin{equation}
\begin{split}
&N_{(\emptyset ,\dots, \emptyset , B)}^\mathrm{Toda}=
\left\langle
j^{(\beta)}_{B} (p_k^{(n)})  \right\rangle=  \\
&=\frac{[ N_n \beta ]_B}{\sqrt{G_{B,B}(0)G_{B,B}(1-\beta)}  }
\,\times
\prod_{a=1}^n
\frac{(-1)^{|B|} [u_{n-a+1}+\cdots+u_n+N_n \beta +a-a \beta]_{B}}
{G_{B,\emptyset}(-(v_{n-a+1}+u_{n-a+1}+\cdots+u_n+N_n\beta+N_{n-a+1}\beta-N_{n-a}\beta+a-a\beta))}\,.
\end{split}
\end{equation}
This is equivalent to (\ref{n1jacki}),
with the identifications of parameters
(where we have omitted the lower index"+" in $v_{a+}$ and $u_{a+}$)
\footnote{There is some degree of freedom to choose the possible identifications.}
\begin{equation}
\begin{split}
\label{id4}
& \mu_{n+1} + a_{n+1} = -N_n\beta \;,\\
&\vdots \\
& \mu_s + a_{n+1} = -\big(u_{s}+\dots +u_n + N_n\beta + (n-s+1)(1 - \beta)\big )\;,\\
&\vdots \\
& \mu_1 + a_{n+1} = -\big(u_{1}+\dots +u_n + N_n\beta + n(1 - \beta)\big )\;,\\
& a_n - a_{n+1} = v_{n} + u_{n} + 2 N_n\beta  - N_{n-1}\beta +1 - \beta \;,\\
&\vdots \\
& a_s - a_{n+1} = v_{s}+u_{s}+\dots +u_{n} + N_n\beta + N_s\beta - N_{s-1}\beta+ (n-s+1)(1-\beta) \;,\\
&\vdots \\
& a_1 - a_{n+1} = v_{1}+u_{1}+\dots +u_{n} + N_n\beta + N_1\beta + n(1-\beta)\;,
\end{split}
\end{equation}
with the restriction $v_2=\cdots=v_{n}=0$ and $v_1=v$.
While this looks complicated, it is simplified in the vector notation
in $\mathbf{R}^{n+1}$,
\begin{equation}
a=\alpha_1+\alpha_2+\beta\sum_a N_a  e_a
+(1-\beta)\rho,\quad
\mu =-\alpha_1-(1-\beta) \rho\,,
\label{param}
\end{equation}
where $ a=\sum_{i=1}^{N+1} a_i  h_i$
and $\mu=\sum_{i=1}^{N+1} \mu_i  h_i$.
We note that $a$ thus written can be identified with
the momentum of the vertex in the intermediate channel.
This gives (\ref{e:id}).
Eq.(\ref{param}) is the desired identification of parameters
in $SU(N+1)$ AGT conjecture \cite{0907.2189,0908.2569}.
We note that this holds for arbitrary $\beta$.

\subsection{General case: arbitrary $\vec Y$, $\beta=1$}
By interpolation method, we have derived that the $(N+1)$-Schur
Selberg integral has the form of
(\ref{nschur}):\\
At $\beta=1$,
{\small
\begin{equation}
\begin{split}
\label{nschurs}
&N_{\vec Y +}^\mathrm{Toda} \\
&=
\left\langle
\chi_{Y_1} (-p_k^{(1)} - (v_{1}+\dots + v_{n}))\dots
\chi_{Y_r} (p_k^{(r-1)} -p_k^{(r)} - \frac{v_{r}+\dots + v_{n}}{\beta})\dots
\chi_{Y_{n+1}} (p_k^{(n)})
 \right\rangle^{SU(n+1)}_{\vec u,\vec v,\beta}
\\
&=\prod_{s=1}^{n}
\bigg\{
(-1)^{|Y_s|}\; \times
\frac{[v_{s}+ N_s - N_{s-1} ]_{Y'_s} }
{[ N_s + N_{s-1}  ]_{Y'_s} }
 \; \times \!
\prod_{1\leq i<j\leq N_{s-1} + N_s}
\frac{(j-i+1)_{Y'_{si}-Y'_{sj}}}{(j-i)_{Y'_{si}-Y'_{sj}}}
\bigg\}
\quad\times
\prod_{1\leq i<j\leq N_{n}}
\frac{(j-i+1)_{Y_{(n+1)i}-Y_{(n+1)j}}}{(j-i)_{Y_{(n+1)i}-Y_{(n+1)j}}}
\\
&\times
\prod_{1\leq t<s\leq n+1}
\bigg\{
\frac{[ v_{t}+u_{t}+\dots +u_{s-1} + N_t- N_{t-1}   ]_{Y'_t} }
{[ v_{t}- v_{s}+u_{t}+\dots +u_{s-1}+ N_{t} - N_{t-1} - N_s  ]_{Y'_t} }
\times
\frac{[-v_{s}+u_{t}+\dots +u_{s-1} - N_s + N_{s-1} ]_{Y_s} }
{[ v_{t}- v_{s}+u_{t}+\dots +u_{s-1} - N_{t-1} - N_s + N_{s-1}  ]_{Y_s} }
 \\
&\qquad\times
\prod_{i=1}^{N_{t}}\prod_{j=1}^{N_{s-1}}
\frac{  v_{t}- v_{s}+u_{t}+\dots +u_{s-1} + N_t- N_{t-1} - N_s + N_{s-1}  +1 -(i +j)}
{ v_{t}- v_{s}+u_{t}+\dots +u_{s-1} + N_t- N_{t-1} - N_s + N_{s-1}   +1 + Y'_{ti} + Y_{sj} -(i +j) }
\bigg\} \;.
\end{split}
\end{equation}
}
Then with the lemmas (\ref{trick1}) to (\ref{trick3})
introduced in the last section and a new assistant
(which only holds at $\beta = 1$),\footnote{Check the appendix for the proof.}
\begin{lemma}
\begin{equation}
\begin{split}
\label{trick4}
\prod_{i=1}^{N_{1}}\prod_{j=1}^{N_{2}} \frac{\big(x+ 1 -(i +j)\beta \big)_{\beta}}
{\big(x+ 1 + A'_i + B_j -(i +j)\beta \big)_{\beta}}
=
\frac{(-1)^{\lvert B \rvert }[x - N_2\beta +1-\beta]_{A'} [x - N_1\beta+1-\beta]_B}
{G_{A,B}(x) G_{B,A}(-x )}
\end{split}
\end{equation}
\end{lemma}

Equation (\ref{nschurs}) transforms to
\begin{equation}
\begin{split}
\label{nsjacks}
&\left\langle
\chi_{Y_1} (-p_k^{(1)} - (v_{1}+\dots + v_{n}))\dots
\chi_{Y_r} (p_k^{(r-1)} -p_k^{(r)} - \frac{v_{r}+\dots + v_{n}}{\beta})\dots
\chi_{Y_{n+1}} (p_k^{(n)})
 \right\rangle^{SU(n+1)}_{\vec u,\vec v,\beta}
 \\
&=\prod_{s=1}^{n}
\bigg\{
(-1)^{|Y_s|} \times\frac{[v_{s}+ N_s - N_{s-1}  ]_{Y'_s}}{G_{Y'_{s},Y'_{s}}(0)}
\bigg\}
\; \times
\frac{[N_n ]_{Y_{n+1}} }{G_{Y_{n+1},Y_{n+1}}(0)} \; \times
\\
&\times
\prod_{1\leq t<s\leq n+1}
\bigg\{
\frac{[v_{t}+u_{t}+\dots +u_{s-1} + N_t- N_{t-1}]_{Y'_t} }
{1 }
\quad\times
\frac{[-\big(v_{s}-u_{t}-\dots -u_{s-1} + N_s - N_{s-1}\big)]_{Y_s} }
{ 1 } \times
\\
&\qquad \times\frac{1}
{ G_{Y_t,Y_s}\big(v_{t}- v_{s}+u_{t}+\dots +u_{s-1} + N_t- N_{t-1} - N_s + N_{s-1}\big) }\; \times
\\
&\qquad \times
\frac{(-1)^{|Y_s|}}
{G_{Y_s,Y_t}\big(-( v_{t}- v_{s}+u_{t}+\dots +u_{s-1} + N_t- N_{t-1} - N_s + N_{s-1})\big)}
\bigg\} \;.
\end{split}
\end{equation}
Further notice that for $\beta = 1$,
\begin{equation}
\label{trick6}
[x]_{A'} = (-1)^{|A|} [-x]_{A} = f_A(x) ,\quad G_{A',A'}(x) = G_{A,A}(x)
\end{equation}
(\ref{nsjacks}) is equivalent to its Nekrasov counterpart (\ref{ninst})
$N_{\vec Y +}^\mathrm{inst}$ at $\beta=1$
with the identifications(\ref{id4})
and the following (where we have again omitted the lower index"+" in $v_{a+}$and $u_{a+}$)
\begin{equation}
\begin{split}
\label{id5}
& a_t - a_{s} =v_{t}- v_{s}+u_{t}+\dots +u_{s-1} + N_t- N_{t-1} - N_s + N_{s-1} \;,\\
& \mu_s + a_{t} = v_{t}+u_{t}+\dots +u_{s-1} + N_t- N_{t-1} \;,\\
& \mu_t + a_{s} = v_{s}-u_{t}-\dots -u_{s-1} + N_s - N_{s-1}\;,\\
& \mu_s + a_{s} = v_{s} + N_s - N_{s-1}\;,
\end{split}
\end{equation}
where $1\leq t<s\leq n$. The above are of course
in accordance with (\ref{id4}) and (\ref{param}).
This implies AGT relation for $SU(n+1)$ at $\beta=1$.

\section{Conclusion and further prospects}
In this paper, we conjectures some formulae for $A_n$ Selberg average
with $n+1$ Jack polynomials and proves AGT relation for
$SU(n+1)$ based on this conjecture.
For the particular combination of Young diagram, namely
$\vec Y=(\emptyset,\cdots,\emptyset,B)$, our proof is
exact since the corresponding Selberg average is already
proved.  For this particular case, the proof is exact for
arbitrary $\beta$.  Our proof is based on a few lemmas
and some of which seem not very straightforward.

The obvious problem is that our formulae for Selberg average
are not based on the explicit evaluation but determined only
by consistency.  So, we need substantial work in the future
to prove them.  One idea may be to use the recursion formula
of $W_{1+\infty}$ algebra \cite{Kanno:2011qv}.  This idea looks
natural since Schur polynomial has simple transformation law
with $W_{1+\infty}$ transformation.  This should work at least
for $\beta=1$.

The difficulty of the proof for $\beta\neq 1$ case has different origin.
For $\beta=1$, we need to compare the factors of
factorized form of $N^\mathrm{Toda}(\vec Y)$
or $N^\mathrm{inst}(\vec Y)$ for each $\vec Y$.  On the other hand,
for $\beta\neq 1$, each factor does not coincide but we need to
compare the sum $\sum_{|\vec Y|=m} N^{\cdots}(\vec Y)$ for
 arbitrary $m=1,2,3,\cdots$
in both side.  This will be certainly more difficult to prove it.
We hope to say something meaningful in such direction,
possibly with the help of the relation with the integrable models.

\section*{Acknowledgements}
The authors would like to thank S. Kanno and S. Shiba
for comments.  YM is partially supported by Grant-in-Aid (\#20540253).
HZ is partially supported by Global COE Program ¡°the Physical Sciences Frontier¡±, MEXT, Japan.

\appendix
\section{Young diagrams}
\begin{figure}[htbp]
\centering
\includegraphics{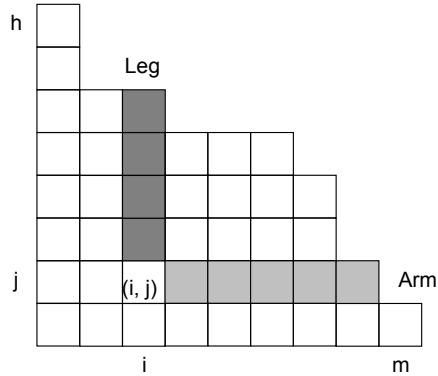}
\caption{example of Young tableaux}
\label{fig:Young}
\end{figure}

Young diagrams are very useful in representing conjugacy classes in group theory.  The above is a Young diagram Y of (8,6,6,5,5,5,4,2,1).
The $i$th column is named as $Y_i$. $h = Y_1$ is the height of Y, while $m = Y'_1$ is called the length of Y,
where $Y'$ stands for the transposed Young diagram.

The arm-length and leg-length of the cell $(i,j)$ in the tableaux $Y$ are denoted by ${\rm Arm}_Y(i,j)$ and ${\rm Leg}_Y(i,j)$ defined separately as
\begin{align}
\label{armleg}
{\rm Arm}_Y(i,j) = Y^{\prime}_j - i, \ \ \ \ {\rm Leg}_Y(i,j) = Y_i - j \;.
\end{align}
For the cell $(i,j)$ = $(3,2)$, the arm-length and leg-length are 5 and 4,
  respectively.

\section{Jack polynomials}\label{a:Jack}
Jack polynomials $J^{(\beta)}_Y[z_1,\cdots,z_M]$ are a kind of symmetric polynomials
of variables $z_1,\cdots, z_M$ labeled by a Young diagram $Y$.
Detailed properties of Jack polynomial is given in \cite{Stanley}.
they are characterized by the fact that they are
the eigenfunctions of Calogero-Sutherland Hamiltonian
written in the form,
\begin{equation}\label{e:CS}
\mathcal{H}=\sum_{i=1}^M D_i^2 +\beta \sum_{i<h} \frac{z_i+z_j}{z_i-z_j}
(D_i-D_j),\quad D_i :=z_i \frac{\partial}{\partial z_i}\,.
\end{equation}
Sometimes they are written as functions of power sum
$p_k(z) = \sum_i z_i^k$.  In the text, we write the Jack polynomial
in terms of them,
$J^{(\beta)}_Y(p_1, p_2,\cdots)\equiv J^{(\beta)}_Y(p_k):=J^{(\beta)}_Y[z_1,\cdots, z_M]$.
The explicit form of low level ones are listed below;
\begin{eqnarray}
\label{jp}
&& J_{[1]}^{(\beta)}(p_k) = p_{1}\nonumber\;,\\
&& J_{[2]}^{(\beta)}(p_k) =  \dfrac{p_2 + \beta p_{1}^2}{\beta + 1}, \quad
 J_{[11]}^{(\beta)}(p_k) =  \dfrac{1}{2} \big( p_{1}^2 - p_2 \big) \;,\\
&&
J_{[3]}^{(\beta)}(p_k) = \dfrac{2 p_3 + 3 \beta p_{1} p_2 + \beta^2 p_{1}^3}{(\beta + 1)(\beta + 2)}, \  J^{(\beta)}_{[21]}(p_k) = \dfrac{(1-\beta) p_{1} p_2 - p_3 + \beta p_{1}^3 }{(\beta + 1)(\beta + 2)}, \
J_{[111]}^{(\beta)}(p_k) = \dfrac{1}{6} p_{1}^3 - \dfrac{1}{2} p_{1} p_2 + \dfrac{1}{3} p_3 \;.\nonumber
\end{eqnarray}
Jack polynomials are orthogonal with each other
$(J_{Y_1}, J_{Y_2})\propto \delta_{Y_1 Y_2}$.
There are two inner products defined for the symmetric polynomial which
has such property.  One is defined in terms of products of power sum,
\begin{equation}
\langle p_1^{k_1}\cdots p_n^{\ell_n}, p_1^{\ell_1}\cdots p_n^{\ell_n}\rangle_\beta
=\delta_{\vec k,\vec\ell} \beta^{-\sum_i k_i} \prod_{i=1}^n i^{k_i} k_i!\,.
\end{equation}
We write the norm for this inner product as $\langle J_Y, J_Y\rangle_\beta
=||J_Y||^2$.
The explicit form of the norm is given in the literature\cite{Stanley,1011.3481}
\begin{align}
||J_A^{(\beta)}||^2 = \dfrac{Q_Y}{P_Y} \;,
\end{align}
with $P_Y$ and $Q_Y$ given by
\begin{align}
P_Y = \prod\limits_{(i,j) \in Y} \Big( \beta ( Y^{\prime}_j - i) + (Y_i - j) + \beta \Big)
=G_{Y,Y}(0)\;,
\end{align}
\begin{align}
Q_Y = \prod\limits_{(i,j) \in Y} \Big( \beta ( Y^{\prime}_j - i) + (Y_i - j) + 1 \Big)
=G_{Y,Y}(1-\beta)\;.
\end{align}
In this paper, we denote the
normalized Jack polynomials as,
\be
\label{jj}
j_Y^{(\beta)}(p):=\frac{J_Y^{(\beta)}(p)}{||J_Y^{(\beta)}||}=\sqrt{\frac{G_{Y,Y}(0)}{G_{Y,Y}(1-\beta)}}J_Y^{(\beta)}(p)\;.
\ee
Especially, at $\b=1$, Jack polynomials reduce to Schur polynomials $\chi_Y$ :
\begin{equation}
j^{(\beta)}_Y|_{\beta=1}=J^{(\beta)}_Y|_{\beta=1}=\chi_Y \;.
\end{equation}
The relation between Jack polynomial and Toda theory is that
Jack polynomial is characterized as the null states of W-algebra, as
discussed, for example, in \cite{AMOS}.  In particular, the
Calogero-Sutherland Hamiltonian (\ref{e:CS}) is written in in terms
of Virasoro and W-generators (see, for example,  eq.(52) of \cite{AMOS}).

The relevance of Jack polynomial in Selberg integral is through
the  Cauchy-Riemann relations,
\begin{eqnarray}
\prod_{i,j} (1-x_i y_j)^{-\beta}= \sum_Y J_Y^{(\beta)}[x]J_Y^{(\beta)}[y]||J_Y||^{-2},\quad
\prod_{i, j}(1+ x_i y_j)=\sum_Y J_{Y'}^{(1/\beta)}[x] J_Y^{(\beta)}[y]\,.
\label{eCR}
\end{eqnarray}
The first property was essentially used in the text.

\section{Formula for general $\beta$}
\label{a:nJack}
Here we write a formula of $A_n$ Selberg average for product of
$n+1$ Jack polynomials which generalizes (\ref{nschur}).
While some modifications on the terms proportional to
$1-\beta$ are required to meet the constraints (\ref{conscheck}), it survives other constraints
which are quite nontrivial.  We write this formula since it may give a useful
hints in the future development, though some modifications are necessary.

The formula for $n+1$ Jack polynomials should be close to the following,
{\small
\begin{equation}
\begin{split}
\label{njackss}
&\left\langle
J^{(\beta)}_{Y_1} (-p_k^{(1)} - \frac{v_{1}+\dots + v_{n}}{\beta})\dots
J^{(\beta)}_{Y_r} (p_k^{(r-1)} -p_k^{(r)} - \frac{v_{r}+\dots + v_{n}}{\beta})\dots
J^{(\beta)}_{Y_{n+1}} (p_k^{(n)})
 \right\rangle^{SU(n+1)}
\\
&=\prod_{s=1}^{n}
\bigg\{
(-1)^{|Y_s|}
\frac{[v_{s}+ N_s\beta - N_{s-1}\beta ]_{Y'_s} }
{[ N_s\beta + N_{s-1}\beta  ]_{Y'_s} }
\prod_{1\leq i<j\leq N_{s-1} + N_s}
\frac{((j-i+1)\beta)_{Y'_{si}-Y'_{sj}}}{((j-i)\beta)_{Y'_{si}-Y'_{sj}}}
\bigg\}
\!\times  \!
\prod_{1\leq i<j\leq N_{n}}
\frac{((j-i+1)\beta)_{Y_{(n+1)i}-Y_{(n+1)j}}}{((j-i)\beta)_{Y_{(n+1)i}-Y_{(n+1)j}}}
\\
&
\times
\prod_{1\leq t<s\leq n}
\bigg\{
\frac{[ v_{t}+u_{t}+\dots +u_{s-1} + N_t\beta- N_{t-1}\beta + (s-t+1)(1 - \beta)  ]_{Y'_t} }
{[ v_{t}- v_{s}+u_{t}+\dots +u_{s-1}+ N_{t}\beta - N_{t-1}\beta - N_s\beta + (s-t+1)(1-\beta) ]_{Y'_t} }
\quad \times \\
&\quad\times
\frac{[-v_{s}+u_{t}+\dots +u_{s-1} - N_s\beta + N_{s-1}\beta + (s-t)(1 - \beta)  ]_{Y_s} }
{[ v_{t}- v_{s}+u_{t}+\dots +u_{s-1} - N_{t-1}\beta - N_s\beta + N_{s-1}\beta + (s-t+1)(1-\beta) ]_{Y_s} }
\quad \times\\
& \times
\prod_{i=1}^{N_{t}} \! \prod_{j=1}^{N_{s-1}} \!\!
\frac{\big(  v_{t}- v_{s}+u_{t}+\dots +u_{s-1} + N_t\beta- N_{t-1}\beta - N_s\beta + N_{s-1}\beta+ (s-t)(1-\beta)  +1 -(i +j)\beta \big)_{\beta}}
{\big(  v_{t}- v_{s}+u_{t}+\dots +u_{s-1} + N_t\beta- N_{t-1}\beta - N_s\beta + N_{s-1}\beta+ (s-t)(1-\beta)   +1 + Y'_{ti} + Y_{sj} -(i +j)\beta \big)_{\beta}}\!\bigg\}\\
&\times
\prod_{1\leq s\leq n}
\bigg\{
\frac{[u_{s}+\dots +u_n + N_n\beta + (n-s+1)(1 - \beta)]_{Y_{n+1}} }
{[v_{s}+u_{s}+\dots +u_{n} + N_n\beta - N_{s-1}\beta+ (n-s+2)(1-\beta)]_{Y_{n+1}} }
\quad \times\\
&\times
\prod_{i=1}^{N_{s}}\prod_{j=1}^{N_{n}}
 \frac{\big( v_{s}+u_{s}+\dots +u_{n} + N_n\beta + N_s\beta - N_{s-1}\beta+ (n-s+1)(1-\beta)   +1 -(i +j)\beta \big)_{\beta}}
{\big(  v_{s}+u_{s}+\dots +u_{n} + N_n\beta + N_s\beta - N_{s-1}\beta+ (n-s+1)(1-\beta)   +1 + Y'_{si} + Y_{(n+1)j} -(i +j)\beta \big)_{\beta}}
\bigg\} \quad.
\end{split}
\end{equation}}
It satisfies consistency conditions with the known results:
\begin{itemize}
\item[(a)] For $Y_{1} = \dots = Y_{n} = \emptyset$, and $Y_{n+1} = B$,
with the help of (\ref{trick5}) the above reduce to the $A_n$
one Jack integral (\ref{n1jacks}).  The proof of this statement is obvious.

\item[(b)] For $n= 1$, $Y_{1} = A$ and $Y_{2} = B$,
 the above reduce to
\begin{eqnarray}
&& \Big< J^{(\beta)}_A(-p_k^{(1)} - \frac{v_{1}}{\beta}) J^{(\beta)}_B(p_k^{(1)}) \Big>^{{SU(2)}}_{u,v,\beta}
= (-1)^{|A|}\times  \dfrac{[v+N\beta]_{A'} [u+N\beta+1-\beta]_B}{[N \beta]_{A'}
[u+v+N\beta+2-2\beta]_B} \times \nonumber\\
&&~~~~~~~~\times
\prod_{1\leq i<j\leq N}\frac{(A'_i-A'_j+(j-i)\beta)_{\beta}}
{((j-i)\beta)_{\beta}}
\prod_{1\leq i<j\leq N}\frac{(B_i-B_j+(j-i)\beta)_{\beta}}
{((j-i)\beta)_{\beta}}
\label{2jackn2}
 \\
&& ~~~~~~~~\times
\prod_{i,j=1}^{ N} \frac{\big(u+v+ 2N\beta+1-\beta +1-(i +j)\beta \big)_{\beta}}
{\big(u+v+ 2N\beta  +A'_i + B_j+1-\beta +1 -(i +j)\beta \big)_{\beta}} \;,
\nonumber
\end{eqnarray}
which is consistent with the $A_1$
two Jack integral (\ref{2jack}) by considering
\begin{align}
j^{(\beta)}_{A}(-p/\beta) = (-1)^{|A|} j^{(1/\beta)}_{A^\prime}(p) \;.
\end{align}

\item[(c)] For $n= 2$, $Y_{1} = R$, $Y_{2} = \emptyset$, and $Y_{3} = B$,
with the help of (\ref{trick5}) the above reduce to
\begin{equation}
\begin{split}
&\left\langle
J^{(\beta)}_{R} (-p_k^{(1)} - \frac{v_{1+}+ v_{(2)+}}{\beta})
J^{(\beta)}_{B} (p_k^{(2)})
\right\rangle_{+}^{SU(3)} \\
&=(-1)^{|R|}\times
\prod_{1\leq i<j\leq N_{1} }
\frac{((j-i+1)\beta)_{R'_{i}-R'_{j}}}{((j-i)\beta)_{R'_{i}-R'_{j}}} \; \times
\prod_{1\leq i<j\leq N_{2}}
\frac{((j-i+1)\beta)_{B_{i}-B_{j}}}{((j-i)\beta)_{B_{i}-B_{j}}}
\\
&\times
\frac{1}
{[ v_{1}- v_{2}+u_{1}+ 2N_{1}\beta - N_{2}\beta + 2(1-\beta) ]_{R'} }  \\
 &\quad\times
\frac{[ v_{1}+u_{1} + N_1\beta +2 - 2\beta  ]_{R'} }
{1}
\quad\times
1
\\
&\times
\prod_{i=1}^{N_{1}}\prod_{j=1}^{N_{2}}
 \frac{\big( v_{1}+u_{1}+u_{2} + N_2\beta + N_1\beta + 2(1-\beta)   +1 -(i +j)\beta \big)_{\beta}}
{\big(  v_{1}+u_{1}+u_{2} + N_2\beta + N_1\beta + 2(1-\beta)   +1 + R'_{i} + B_{j} -(i +j)\beta \big)_{\beta}} \\
&\quad\times
\frac{[u_{1}+u_2 + N_2\beta + 2(1 - \beta)]_{B} }
{[v_{1}+u_{1} +u_{2} + N_2\beta + 3(1-\beta)]_{B} }
\quad\times
\frac{[v_{1}+ N_1\beta ]_{R'} }
{[ N_1\beta ]_{R'} }
 \\
 &\times
\frac{1 }
{[v_{2}+u_{2} + 2N_2\beta -N_1\beta + 2(1-\beta)]_{B} } \\
&\quad\times
\frac{[u_2 + N_2\beta + (1 - \beta)]_{B} }
{1 }
\quad\times
1 \;.
\end{split}
\end{equation}
Notice the shift in $j_{R}$'s argument, and the restrictions $v_2=0,\quad v_1=v$, $\quad v_1 +v_2= \beta -1$ (this last restrictions is only claimed by Warnaar's $A_2$ two Jack integral), the above is
consistent with the $A_2$
two Jack integral given by Warnaar\cite{0901.4176} as below
\begin{equation}
\begin{split}
\label{2jacka2}
& \Big< J^{(\beta)}_R(p_k^{(1)}) J^{(\beta)}_B(p_k^{(2)}) \Big>^{{SU(3)}}_{u,v,\beta} \\
&=
\prod_{1\leq i<j\leq N_1}\frac{((j-i+1)\beta)_{R_i-R_j}}
{((j-i)\beta)_{R_i-R_j}}
\prod_{1\leq i<j\leq N_2}\frac{((j-i+1)\beta)_{B_i-B_j}}
{((j-i)\beta)_{B_i-B_j}}
 \\
&\times
\  \dfrac{[u_1 +N_1\beta +1-\beta]_{R}  }
{[v_1+u_1 +2N_1\beta -N_2\beta +2-2\beta]_{R} } \times
\dfrac{ \quad [u_2 +N_2\beta +1-\beta]_{B} }
{[v_2+u_2 +2N_2\beta -N_1\beta +2-2\beta]_{B}} \times
\nonumber\\
&\times
\prod_{i=1}^{ N_1}\prod_{j=1}^{ N_2}
\frac{\big(u_1+u_2+ N_1\beta+N_2\beta+1-\beta +1-(i +j)\beta \big)_{\beta}}
{\big(u_1+u_2+ N_1\beta+N_2\beta  +R_i + B_j+1-\beta +1 -(i +j)\beta \big)_{\beta}} \;.
\nonumber
\end{split}
\end{equation}

\item[(d)] For $N_n =0$ (so that $u_n =v_n=0$, and $Y_{n+1} = \emptyset$), the above reduce to
{\small
\begin{equation}
\begin{split}
&\left\langle
J^{(\beta)}_{Y_1} (-p_k^{(1)} - \frac{v_{1}+\dots + v_{n}}{\beta})\dots
J^{(\beta)}_{Y_r} (p_k^{(r-1)} -p_k^{(r)} - \frac{v_{r}+\dots + v_{n}}{\beta})\dots
J^{(\beta)}_{Y_{n+1}} (p_k^{(n)})
 \right\rangle^{SU(n+1)}
\\
&=\prod_{s=1}^{n-1}
\bigg\{
(-1)^{|Y_s|}
\frac{[v_{s}+ N_s\beta - N_{s-1}\beta ]_{Y'_s} }
{[ N_s\beta + N_{s-1}\beta  ]_{Y'_s} }
\prod_{1\leq i<j\leq N_{s-1} + N_s}
\frac{((j-i+1)\beta)_{Y'_{si}-Y'_{sj}}}{((j-i)\beta)_{Y'_{si}-Y'_{sj}}}
\bigg\}
\!\times  \!
\prod_{1\leq i<j\leq N_{n-1}}
\frac{((j-i+1)\beta)_{Y_{(n)i}-Y_{(n)j}}}{((j-i)\beta)_{Y_{(n)i}-Y_{(n)j}}}
\\
&
\times
\prod_{1\leq t<s\leq n-1}
\bigg\{
\frac{[ v_{t}+u_{t}+\dots +u_{s-1} + N_t\beta- N_{t-1}\beta + (s-t+1)(1 - \beta)  ]_{Y'_t} }
{[ v_{t}- v_{s}+u_{t}+\dots +u_{s-1}+ N_{t}\beta - N_{t-1}\beta - N_s\beta + (s-t+1)(1-\beta) ]_{Y'_t} }
\quad \times \\
&\quad\times
\frac{[-v_{s}+u_{t}+\dots +u_{s-1} - N_s\beta + N_{s-1}\beta + (s-t)(1 - \beta)  ]_{Y_s} }
{[ v_{t}- v_{s}+u_{t}+\dots +u_{s-1} - N_{t-1}\beta - N_s\beta + N_{s-1}\beta + (s-t+1)(1-\beta) ]_{Y_s} }
\quad \times\\
& \times
\prod_{i=1}^{N_{t}} \! \prod_{j=1}^{N_{s-1}} \!\!
\frac{\big(  v_{t}- v_{s}+u_{t}+\dots +u_{s-1} + N_t\beta- N_{t-1}\beta - N_s\beta + N_{s-1}\beta+ (s-t)(1-\beta)  +1 -(i +j)\beta \big)_{\beta}}
{\big(  v_{t}- v_{s}+u_{t}+\dots +u_{s-1} + N_t\beta- N_{t-1}\beta - N_s\beta + N_{s-1}\beta+ (s-t)(1-\beta)   +1 + Y'_{ti} + Y_{sj} -(i +j)\beta \big)_{\beta}}\!\bigg\}\\
&\times
\prod_{1\leq t\leq n-1}
\bigg\{1\times
\frac{[u_{t}+\dots +u_{n-1} + N_{n-1}\beta + (n-s)(1 - \beta)]_{Y_{n}} }
{[v_{t}+u_{t}+\dots +u_{n-1} + N_{n-1}\beta - N_{t-1}\beta+ (n-t+1)(1-\beta)]_{Y_{n}} }
\quad \times\\
&\times
\prod_{i=1}^{N_{t}}\prod_{j=1}^{N_{n-1}}
 \frac{\big( v_{t}+u_{t}+\dots +u_{n-1} + N_{n-1}\beta + N_t\beta - N_{t-1}\beta+ (n-t)(1-\beta)   +1 -(i +j)\beta \big)_{\beta}}
{\big(  v_{t}+u_{t}+\dots +u_{n-1} + N_{n-1}\beta + N_t\beta - N_{t-1}\beta+ (n-t)(1-\beta)   +1 + Y'_{si} + Y_{(n+1)j} -(i +j)\beta \big)_{\beta}}
\bigg\}\\
&\times 1 \; .
\end{split}
\end{equation}
}
This is just the expression of
$$\left\langle
J_{Y_1} (-p_k^{(1)} - \frac{v_{1}+\dots + v_{(n-1)}}{\beta})\dots
J_{Y_r} (p_k^{(r-1)} -p_k^{(r)} - \frac{v_{r}+\dots + v_{(n-1)}}{\beta})\dots
J_{Y_{n}} (p_k^{(n-1)})
 \right\rangle^{A_{n-1}}.$$
\end{itemize}

\section{Proof of consistency relations}
\label{a:consistency}
\begin{figure}[htbp]
\centering
\includegraphics{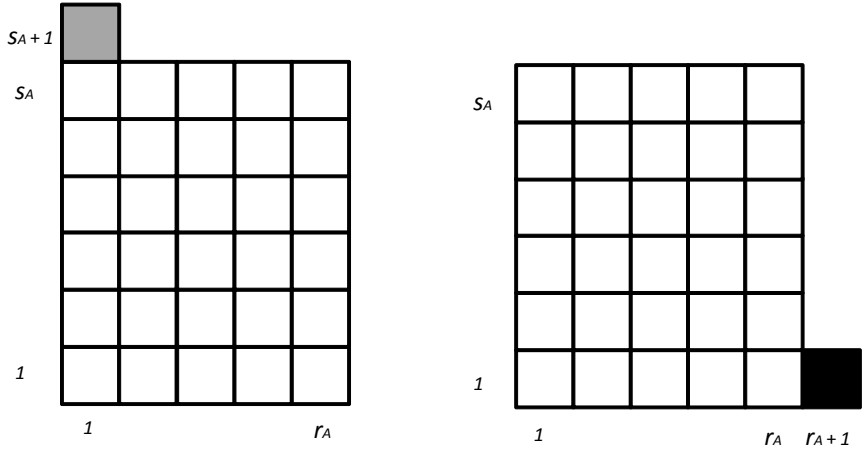}
\caption{The white cells stands for $A$, with length $r_A$ and height $s_A$. the left is the diagram of $\hat{A}$, with an extra grey cell compared to $A$; the right is the diagram of $\breve{A}$, with an extra black cell compared to $A$. $A_i = s_A$, $A'_j = r_A$, $\hat{A}_1 = s_A +1$, and $\breve{A}'_1 = r_A +1$.
 }
\label{fig:0}
\end{figure}

Here we present the detailed computation
of  the second sets of consistency conditions (\ref{conscheck}) in the text.\\
When $n=2$ ($SU(3)$ case), making use of (\ref{trick6}), and setting $Y_1=R$, $Y_2=A$, $Y_3=B$,
 the conjecture (\ref{nschur}) becomes
\begin{equation}
\begin{split}
&\left\langle
\chi_{R} (-p_k^{(1)} - v)
\chi_{A} (p_k^{(1)} -p_k^{(2)})
\chi_{B} (p_k^{(2)})
\right\rangle^{SU(3)}
\\
&=
\frac{[-v_{1}-N_1 ]_{R}}{G_{R,R}(0) } \; \times
\frac{[N_1 -N_2 ]_{A} }{G_{A,A}(0) } \; \times
\frac{[N_2 ]_{B}}{G_{B,B}(0) }
\\
&\quad\times
\frac{[ -v-u_{1} - N_1 ]_{R} }
{[ -v-u_{1} - N_1 + N_{2} ]_{R} }
\quad\times
\frac{[u_{1}+ N_{1}- N_2 ]_{A} }
{[ v + u_{1}+ N_{1}- N_2 ]_{A} }
\quad\times
\frac{[u_{1}+u_2 + N_2]_{B} }
{[v+u_{1} +u_{2} + N_2]_{B} }
\quad\times
\frac{[u_2 + N_2]_{B} }
{[u_{2}-N_1 + N_2]_{B} }
\\
&\quad\times
\prod_{j=1}^{N_{1}}\prod_{i=1}^{N_{1}}
\frac{ v+u_{1} + 2N_1 - N_2  +1 -(i +j)}
{ v+u_{1} + 2N_1 - N_2  +1
 + R'_{j} + A_{i} -(i +j)}\\
&\quad\times
\prod_{j=1}^{N_{1}}\prod_{i=1}^{N_{2}}
 \frac{ v+u_{1}+u_{2} + N_1+ N_2  +1 -(i +j)}
{  v +u_{1}+u_{2}+ N_1+ N_2  +1 + R'_{j} + B_{i} -(i +j)} \\
 &\quad\times
\prod_{j=1}^{N_{2}}\prod_{i=1}^{N_{2}}
 \frac{ u_{2}- N_1 + 2N_2 +1 -(i +j) }
{u_{2}- N_1 + 2N_2   +1 + A'_{j} + B_{i} -(i +j) }\quad ,
\end{split}
\end{equation}
where we have switched the name of $i$ and $j$ in the last three lines.
\\
\\
For simplicity, we consider the case with $R, A, B$ being rectangle Young diagrams, when (\ref{schuid}) reduce to
\begin{equation}
p_1 \chi_A(p_k)=\chi_{\hat{A}}(p_k) +\chi_{\breve{A}}(p_k)\;,
\end{equation}
as illustrated in Figure \ref{fig:0}.

Now at $\beta=1$, there are
\begin{equation}
[ x ]_A
=\prod_{i=1}^{r_A}\prod_{j=1}^{s_A}( x -  i +  j )\quad, \quad
G_{A,A}(0) = \prod_{i=1}^{r_A}\prod_{j=1}^{s_A} (r_A +s_A -i-j+1)\quad.
\end{equation}
Furthermore with the information given in Figure \ref{fig:0}, we find several lemmas shown below
\begin{equation}
\frac{[ x ]_{\hat{A}}}{[ x ]_A}
=x +s_A
\quad, \quad
\frac{[ x ]_{\breve{A}}}{[ x ]_A}
=x -r_A \quad,
\end{equation}

\begin{equation}
\frac{G_{A,A}(0)}{G_{\hat{A},\hat{A}}(0)}
=\prod_{j=1}^{s_A} \frac{r_A +s_A -j}{r_A +s_A-j+1} =\frac{r_A}{r_A + s_A} \quad, \quad
\frac{G_{A,A}(0)}{G_{\breve{A},\breve{A}}(0)}
=\prod_{i=1}^{r_A} \frac{r_A +s_A -i}{r_A +s_A-i+1} =\frac{s_A}{r_A + s_A}\quad,
\end{equation}

\begin{equation}
\begin{split}
&\prod_{j=1}^{N_{1}}\prod_{i=1}^{N_{2}}
 \frac{ x +1 + A'_{j} + B_{i}-(i +j) }
{ x +1 + \hat{A}'_{j} + B_{i}-(i +j) }\\
&=\prod_{i=1}^{N_{2}}
 \frac{ x +1 + 0 + B_{i}-i -(s_A +1) }
{ x +1 + 1 + B_{i}-i -(s_A +1)}=
\prod_{i=1}^{r_B}
 \frac{x +s_B-s_A -i }
{ x +s_B-s_A -i +1}
\times
\prod_{i=r_B +1}^{N_{2}}
 \frac{ x -s_A -i  }
{ x-s_A -i +1 }\\
&
= \frac{ x +s_B-s_A -r_B }
{ x +s_B-s_A}
\times
 \frac{ x -s_A -N_2}
{ x-s_A -r_B }\quad,
\end{split}
\end{equation}

\begin{equation}
\begin{split}
&\prod_{j=1}^{N_{1}}\prod_{i=1}^{N_{2}}
 \frac{ x +1 + A'_{j} + B_{i}-(i +j) }
{ x +1 + \breve{A}'_{j} + B_{i}-(i +j) }\\
&=\prod_{i=1}^{N_{2}}
 \frac{ x +1 + r_A + B_{i}-i-1 }
{ x +1 + r_A +1 + B_{i}-i-1}=
\prod_{i=1}^{r_B}
 \frac{x +s_B+ r_A -i }
{ x +s_B+ r_A -i +1}
\times
\prod_{i=r_B +1}^{N_{2}}
 \frac{ x + r_A -i  }
{ x+ r_A -i +1 }\\
&
= \frac{ x +s_B+ r_A -r_B }
{ x +s_B+ r_A}
\times
 \frac{ x + r_A -N_2}
{ x+ r_A -r_B }\quad,
\end{split}
\end{equation}

\begin{equation}
\begin{split}
&\prod_{j=1}^{N_{1}}\prod_{i=1}^{N_{2}}
 \frac{ x +1 + A'_{j} + B_{i}-(i +j) }
{ x +1 + A'_{j} + \hat{B}_{i}-(i +j) }\\
&=\prod_{j=1}^{N_{1}}
 \frac{ x +1 +A'_{j}+ s_B -1-j }
{ x +1 +A'_{j}+ s_B+1 -1-j}=
\prod_{j=1}^{s_A}
 \frac{x+ r_A +s_B-j }
{ x+ r_A +s_B-j +1}
\times
\prod_{j=s_A +1}^{N_{1}}
 \frac{ x + s_B -j  }
{ x + s_B -j +1 }\\
&
= \frac{ x + r_A +s_B -s_A }
{ x + r_A +s_B}
\times
 \frac{ x + s_B -N_1}
{ x + s_B -s_A }\quad,
\end{split}
\end{equation}
and
\begin{equation}
\begin{split}
&\prod_{j=1}^{N_{1}}\prod_{i=1}^{N_{2}}
 \frac{ x +1 + A'_{j} + B_{i}-(i +j) }
{ x +1 + A'_{j} + \breve{B}_{i}-(i +j) }\\
&=\prod_{j=1}^{N_{1}}
 \frac{ x +1 +A'_{j}+ 0 -(r_B +1)-j }
{ x +1 +A'_{j}+ 1 -(r_B +1)-j }=
\prod_{j=1}^{s_A}
 \frac{x+ r_A -r_B-j }
{ x+ r_A -r_B-j +1}
\times
\prod_{j=s_A +1}^{N_{1}}
 \frac{ x -r_B -j  }
{ x-r_B -j +1 }\\
&
= \frac{ x + r_A -r_B -s_A }
{ x + r_A -r_B}
\times
 \frac{ x -r_B -N_1}
{ x -r_B-s_A }\quad.
\end{split}
\end{equation}
With the help of the above lemmas, we can calculate that
\begin{equation}
\begin{split}
&\frac{\left\langle
\chi_{\hat{R}} (-p_k^{(1)} - v)
\chi_{A} (p_k^{(1)} -p_k^{(2)})
\chi_{B} (p_k^{(2)})
\right\rangle}
{\left\langle
\chi_{R} (-p_k^{(1)} - v)
\chi_{A} (p_k^{(1)} -p_k^{(2)})
\chi_{B} (p_k^{(2)})
\right\rangle}
\\
&=
\frac{[-v_{1}-N_1 ]_{\hat{R}}}{[-v_{1}-N_1 ]_{R} } \; \times
\frac{G_{R,R}(0)}{G_{\hat{R},\hat{R}}(0) } \; \times
\frac{[ -v-u_{1} - N_1 ]_{\hat{R}} }
{[ -v-u_{1} - N_1 ]_{R} }
\times
\frac{[ -v-u_{1} - N_1 + N_{2} ]_{R} }
{[ -v-u_{1} - N_1 + N_{2} ]_{\hat{R}} }
\\
&\quad\times
\prod_{j=1}^{N_{1}}\prod_{i=1}^{N_{1}}
\frac{ v+u_{1} + 2N_1 - N_2  +1 + R'_{j} + A_{i} -(i +j)}
{ v+u_{1} + 2N_1 - N_2  +1 + \hat{R}'_{j} + A_{i} -(i +j)}\\
&\quad\times
\prod_{j=1}^{N_{1}}\prod_{i=1}^{N_{2}}
 \frac{  v +u_{1}+u_{2}+ N_1+ N_2  +1 + R'_{j} + B_{i} -(i +j)}
{  v +u_{1}+u_{2}+ N_1+ N_2  +1 + \hat{R}'_{j} + B_{i} -(i +j)}
\\
&=
(-v-N_1 +s_R)\times \frac{r_R}{r_R +s_R}\times \frac{-v-u_1 -N_1 +s_R}{-v-u_1 -N_1+N_2 +s_R}\times\\
& \quad \times
\frac{v+u_1+2N_1-N_2+s_A-s_R-r_A}{v+u_1+2N_1-N_2+s_A-s_R}\times
\frac{v+u_1+N_1-N_2-s_R}{v+u_1+2N_1-N_2-s_R-r_A} \times \\
& \quad \times
\frac{v+u_1+u_2+N_1+N_2+s_B-s_R-r_B}{v+u_1+u_2+N_1+N_2+s_B-s_R}\times
\frac{v+u_1+u_2+N_1-s_R}{v+u_1+u_2+N_1+N_2-s_R-r_B} \quad.
\end{split}
\end{equation}

Likewise, we have
\begin{equation}
\begin{split}
&\frac{\left\langle
\chi_{\breve{R}} (-p_k^{(1)} - v)
\chi_{A} (p_k^{(1)} -p_k^{(2)})
\chi_{B} (p_k^{(2)})
\right\rangle}
{\left\langle
\chi_{R} (-p_k^{(1)} - v)
\chi_{A} (p_k^{(1)} -p_k^{(2)})
\chi_{B} (p_k^{(2)})
\right\rangle}
\\
&=
(-v-N_1 -r_R)\times \frac{s_R}{r_R +s_R}\times \frac{-v-u_1 -N_1 -r_R}{-v-u_1 -N_1+N_2 -r_R}\times\\
& \quad \times
\frac{v+u_1+2N_1-N_2+s_A+r_R-r_A}{v+u_1+2N_1-N_2+s_A+r_R}\times
\frac{v+u_1+N_1-N_2+r_R}{v+u_1+2N_1-N_2+r_R-r_A} \times \\
& \quad \times
\frac{v+u_1+u_2+N_1+N_2+s_B+r_R-r_B}{v+u_1+u_2+N_1+N_2+s_B+r_R}\times
\frac{v+u_1+u_2+N_1+r_R}{v+u_1+u_2+N_1+N_2+r_R-r_B} \quad,
\end{split}
\end{equation}

\begin{equation}
\begin{split}
&\frac{\left\langle
\chi_{R} (-p_k^{(1)} - v)
\chi_{\hat{A}} (p_k^{(1)} -p_k^{(2)})
\chi_{B} (p_k^{(2)})
\right\rangle}
{\left\langle
\chi_{R} (-p_k^{(1)} - v)
\chi_{A} (p_k^{(1)} -p_k^{(2)})
\chi_{B} (p_k^{(2)})
\right\rangle}
\\
&=
(N_1-N_2 +s_A)\times \frac{r_A}{r_A +s_A}\times \frac{u_1 +N_1-N_2 +s_A}{v+u_1 +N_1-N_2 +s_A}\times\\
& \quad \times
\frac{v+u_1+2N_1-N_2+r_R+s_A-s_R}{v+u_1+2N_1-N_2+r_R+s_A}\times
\frac{v+u_1+N_1-N_2+s_A}{v+u_1+2N_1-N_2+s_A-s_R}\times \\
& \quad \times
\frac{u_2-N_1+2N_2+s_B-s_A-r_B}{u_2-N_1+2N_2+s_B-s_A}\times
\frac{u_2-N_1+N_2-s_A}{u_2-N_1+2N_2-s_A-r_B} \quad,
\end{split}
\end{equation}

\begin{equation}
\begin{split}
&\frac{\left\langle
\chi_{R} (-p_k^{(1)} - v)
\chi_{\breve{A}} (p_k^{(1)} -p_k^{(2)})
\chi_{B} (p_k^{(2)})
\right\rangle}
{\left\langle
\chi_{R} (-p_k^{(1)} - v)
\chi_{A} (p_k^{(1)} -p_k^{(2)})
\chi_{B} (p_k^{(2)})
\right\rangle}
\\
&=
(N_1-N_2 -r_A)\times \frac{s_A}{r_A +s_A}\times \frac{u_1 +N_1-N_2 -r_A}{v+u_1 +N_1-N_2 -r_A}\times\\
& \quad \times
\frac{v+u_1+2N_1-N_2+r_R-r_A-s_R}{v+u_1+2N_1-N_2+r_R-r_A}\times
\frac{v+u_1+N_1-N_2-r_A}{v+u_1+2N_1-N_2-r_A-s_R}\times \\
& \quad \times
\frac{u_2-N_1+2N_2+s_B+r_A-r_B}{u_2-N_1+2N_2+s_B+r_A}\times
\frac{u_2-N_1+N_2+r_A}{u_2-N_1+2N_2+r_A-r_B} \quad,
\end{split}
\end{equation}

\begin{equation}
\begin{split}
&\frac{\left\langle
\chi_{R} (-p_k^{(1)} - v)
\chi_{A} (p_k^{(1)} -p_k^{(2)})
\chi_{\hat{B}} (p_k^{(2)})
\right\rangle}
{\left\langle
\chi_{R} (-p_k^{(1)} - v)
\chi_{A} (p_k^{(1)} -p_k^{(2)})
\chi_{B} (p_k^{(2)})
\right\rangle}
\\
&=
(N_2 +s_B)\times \frac{r_B}{r_B +s_B}\times \frac{u_1 +u_2+N_2 +s_B}{v+u_1 +u_2+N_2 +s_B}\times
\frac{u_2+N_2 +s_B}{u_2-N_1+N_2 +s_B}
\\
& \quad \times
\frac{v+u_1+u_2+N_1+N_2+r_R+s_B-s_R}{v+u_1+u_2+N_1+N_2+r_R+s_B}\times
\frac{v+u_1+u_2+N_2+s_B}{v+u_1+u_2+N_1+N_2+s_B-s_R}\times \\
& \quad \times
\frac{u_2-N_1+2N_2+r_A+s_B-s_A}{u_2-N_1+2N_2+r_A+s_B}\times
\frac{u_2-N_1+N_2+s_B}{u_2-N_1+2N_2+s_B-s_A} \quad,
\end{split}
\end{equation}

and
\begin{equation}
\begin{split}
&\frac{\left\langle
\chi_{R} (-p_k^{(1)} - v)
\chi_{A} (p_k^{(1)} -p_k^{(2)})
\chi_{\breve{B}} (p_k^{(2)})
\right\rangle}
{\left\langle
\chi_{R} (-p_k^{(1)} - v)
\chi_{A} (p_k^{(1)} -p_k^{(2)})
\chi_{B} (p_k^{(2)})
\right\rangle}
\\
&=
(N_2 -r_B)\times \frac{s_B}{r_B +s_B}\times \frac{u_1 +u_2+N_2 -r_B}{v+u_1 +u_2+N_2 -r_B}\times
\frac{u_2+N_2 -r_B}{u_2-N_1+N_2 -r_B}
\\
& \quad \times
\frac{v+u_1+u_2+N_1+N_2+r_R-r_B-s_R}{v+u_1+u_2+N_1+N_2+r_R-r_B}\times
\frac{v+u_1+u_2+N_2-r_B}{v+u_1+u_2+N_1+N_2-r_B-s_R}\times \\
& \quad \times
\frac{u_2-N_1+2N_2+r_A-r_B-s_A}{u_2-N_1+2N_2+r_A-r_B}\times
\frac{u_2-N_1+N_2-r_B}{u_2-N_1+2N_2-r_B-s_A} \quad.
\end{split}
\end{equation}

Summing $v$ and the above six expressions together, we obtain
\begin{equation}
\begin{split}
&v
+\frac{\left\langle
\chi_{\hat{R}} (-p_k^{(1)} - v)
\chi_{A} (p_k^{(1)} -p_k^{(2)})
\chi_{B} (p_k^{(2)})
\right\rangle}
{\left\langle
\chi_{R} (-p_k^{(1)} - v)
\chi_{A} (p_k^{(1)} -p_k^{(2)})
\chi_{B} (p_k^{(2)})
\right\rangle}
+\frac{\left\langle
\chi_{\breve{R}} (-p_k^{(1)} - v)
\chi_{A} (p_k^{(1)} -p_k^{(2)})
\chi_{B} (p_k^{(2)})
\right\rangle}
{\left\langle
\chi_{R} (-p_k^{(1)} - v)
\chi_{A} (p_k^{(1)} -p_k^{(2)})
\chi_{B} (p_k^{(2)})
\right\rangle}\\
&\quad
+\frac{\left\langle
\chi_{R} (-p_k^{(1)} - v)
\chi_{\hat{A}} (p_k^{(1)} -p_k^{(2)})
\chi_{B} (p_k^{(2)})
\right\rangle}
{\left\langle
\chi_{R} (-p_k^{(1)} - v)
\chi_{A} (p_k^{(1)} -p_k^{(2)})
\chi_{B} (p_k^{(2)})
\right\rangle}
+\frac{\left\langle
\chi_{R} (-p_k^{(1)} - v)
\chi_{\breve{A}} (p_k^{(1)} -p_k^{(2)})
\chi_{B} (p_k^{(2)})
\right\rangle}
{\left\langle
\chi_{R} (-p_k^{(1)} - v)
\chi_{A} (p_k^{(1)} -p_k^{(2)})
\chi_{B} (p_k^{(2)})
\right\rangle}\\
&\quad
+\frac{\left\langle
\chi_{R} (-p_k^{(1)} - v)
\chi_{A} (p_k^{(1)} -p_k^{(2)})
\chi_{\hat{B}} (p_k^{(2)})
\right\rangle}
{\left\langle
\chi_{R} (-p_k^{(1)} - v)
\chi_{A} (p_k^{(1)} -p_k^{(2)})
\chi_{B} (p_k^{(2)})
\right\rangle}
\frac{\left\langle
\chi_{R} (-p_k^{(1)} - v)
\chi_{A} (p_k^{(1)} -p_k^{(2)})
\chi_{\breve{B}} (p_k^{(2)})
\right\rangle}
{\left\langle
\chi_{R} (-p_k^{(1)} - v)
\chi_{A} (p_k^{(1)} -p_k^{(2)})
\chi_{B} (p_k^{(2)})
\right\rangle}
=0 \quad.
\end{split}
\end{equation}

This reproduces (\ref{conscheck}), which serves as
a quite nontrivial check of our conjecture (\ref{nschur}).

\section{Proof of the lemmas}
\subsection{Proof of Lemma 1}
\paragraph{Lemma 1}
\begin{equation}
\prod_{1\leq i<j\leq N}\frac{((j-i+1)\beta)_{B_i-B_j}}
{((j-i)\beta)_{B_i-B_j}}=
\frac{[ N \beta ]_B}{G_{B,B}(0) }
\end{equation}

\begin{figure}[htbp]
\centering
\includegraphics{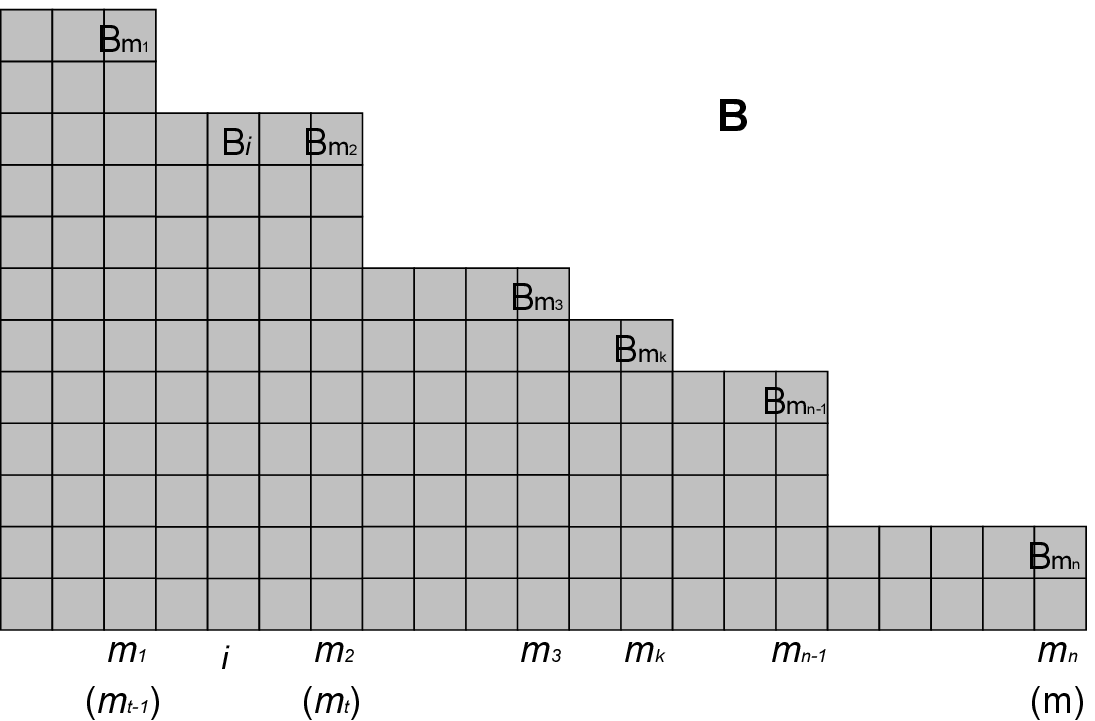}
\caption{ }
\label{fig:2}
\end{figure}

\paragraph{Proof:}
Since
$
(x)_k
= \frac{\Gamma (x + k)} { \Gamma (x )}
$,
we obtain
\begin{equation}
\frac{((j-i+1)\beta)_{B_i-B_j}}
{((j-i)\beta)_{B_i-B_j}}
= \frac{\Gamma ((j-i+1)\beta+B_i-B_j)}
{\Gamma ((j-i+1)\beta)}
\times
\frac{\Gamma ((j-i)\beta)}{\Gamma ((j-i)\beta+B_i-B_j)}
=\frac{(B_i-B_j+(j-i)\beta)_{\beta}}
{((j-i)\beta)_{\beta}} \;.
\end{equation}
So we only need to prove the following
\begin{equation}
\prod_{1\leq i<j\leq N}\frac{(B_i-B_j+(j-i)\beta)_{\beta}}
{((j-i)\beta)_{\beta}}
=
\frac{\prod\limits_{(i,j) \in B} ( N\beta - \beta ( i - 1) +  j - 1 )}
{\prod\limits_{(i,j) \in B} \Big[ \beta (B'_j - i) + (B_i - j) + \beta \Big] }
=
\frac{[ N \beta ]_B}{G_{B,B}(0) }\;.
\end{equation}
Suppose  the length of $B$ to be $m$, The left hand side can be expressed as
\begin{equation}
\prod_{1\leq i<j\leq N}\frac{(B_i-B_j+(j-i)\beta)_{\beta}}
{((j-i)\beta)_{\beta}}
=
\prod_{i=1}^{m}
\prod_{j=m+1}^{N}
\frac{(B_i+(j-i)\beta)_{\beta}}
{((j-i)\beta)_{\beta}}
\times
\prod_{i=1}^{m-1}
\prod_{j=i+1}^{m}
\frac{(B_i-B_j+(j-i)\beta)_{\beta}}
{((j-i)\beta)_{\beta}}\;,
\end{equation}
where
\begin{equation}
\begin{split}
&\prod_{i=1}^{m}
\prod_{j=m+1}^{N}
\frac{(B_i+(j-i)\beta)_{\beta}}
{((j-i)\beta)_{\beta}}
\\
&=
\prod_{i=1}^{m}
\prod_{j=m+1}^{N}
\frac{(1+(j-i)\beta)_{\beta}}
{((j-i)\beta)_{\beta}}
\,
\frac{(2+(j-i)\beta)_{\beta}}
{(1+(j-i)\beta)_{\beta}}
\cdots
\frac{(B_i+(j-i)\beta)_{\beta}}
{(B_i -1 +(j-i)\beta)_{\beta}}
\\
&=
\prod_{i=1}^{m}
\prod_{j=m+1}^{N}
\frac{(j-i+1)\beta}
{(j-i)\beta}
\,
\frac{1+(j-i+1)\beta}
{1+(j-i)\beta}
\cdots
\frac{B_i -1+(j-i+1)\beta}
{B_i -1+(j-i)\beta}
\\
&=
\prod_{i=1}^{m}
\prod_{j=m+1}^{N}
\prod_{k=1}^{B_i}
\frac{k -1+(j-i+1)\beta}
{k -1+(j-i)\beta}
=
\prod_{i=1}^{m}
\prod_{j=1}^{B_i}
\prod_{k=m+1}^{N}
\frac{j -1+(k-i+1)\beta}
{j -1+(k-i)\beta}=
\\
&=
\prod_{(i,j) \in B}
\prod_{k=m+1}^{N}
\frac{j -1+(k-i+1)\beta}
{j -1+(k-i)\beta}
=
\prod_{(i,j) \in B}
\frac{j -1+(N-i+1)\beta}
{j -1+(m-i+1)\beta} \;.
\end{split}
\end{equation}
So what is left is to prove the following equation:
\begin{equation}
\label{lemma1}
\prod_{i=1}^{m-1}
\prod_{j=i+1}^{m}
\frac{(B_i-B_j+(j-i)\beta)_{\beta}}
{((j-i)\beta)_{\beta}}
=
\prod_{i=1}^{m}
\prod_{j=1}^{B_i} \frac{( m\beta - \beta ( i - 1) +  j - 1 )}
{ \Big[ \beta (B'_j - i) + (B_i - j) + \beta \Big] }\;.
\end{equation}
Notice when $1 \leq j \leq B_m$, we have $B'_j= m $,
\begin{equation}
\prod_{j=1}^{B_m} \frac{( m\beta - \beta ( m - 1) +  j - 1 )}
{ \Big[ \beta (m - m) + (B_m - j) + \beta \Big] }
=1\,.
\end{equation}
Thus the sufficient condition of (\ref{lemma1}) is
\begin{equation}
\label{lemma1a}
\prod_{j=i+1}^{m}
\frac{(B_i-B_j+(j-i)\beta)_{\beta}}
{((j-i)\beta)_{\beta}}
=
\prod_{j=1}^{B_i} \frac{( m\beta - \beta ( i - 1) +  j - 1 )}
{ \Big[ \beta (B'_j - i) + (B_i - j) + \beta \Big] } \;,
\end{equation}
which becomes our new goal. \\ \\
In Figure \ref{fig:2}, we have

\[B'_j =\begin{cases}
 m_1 & B_{m_2}+1\leq j\leq B_{m_1}\\
 m_2 & B_{m_3}+1\leq j\leq B_{m_2}\\
\vdots & \vdots  \\
 m_n & 1\leq j\leq B_{m_n}
\end{cases}\]

and if $m_{t-1}+1\leq i \leq m_t$, we have $B_i = B_{m_t}$. Besides, We define $B_{m_{n+1}}=0$. \\ \\
Now the denominator on the right hand side of (\ref{lemma1a}) is
\begin{equation}
\begin{split}
&R_1=
\prod_{j=1}^{B_i}
 \Big[ \beta (B'_j - i) + (B_i - j) + \beta \Big]
 = \prod_{k=t}^{n}
\;
\prod_{j=B_{m_{k+1}}+1}^{B_{m_k}}
 \Big[ (B_i - j)+ \beta (m_k - i +1) \Big]\;,
\end{split}
\end{equation}
and the left hand side of (\ref{lemma1a}) is
\begin{equation}
\begin{split}
&L = \prod_{j=i+1}^{m}
\frac{(B_i-B_j+(j-i)\beta)_{\beta}}
{((j-i)\beta)_{\beta}}
\\
& =
\prod_{j=m_t+1}^{m}
\frac{(B_i-B_j+(j-i)\beta)_{\beta}}
{((j-i)\beta)_{\beta}}
\\
& =
\prod_{j=m_t+1}^{m}
\frac{(j-i+1)\beta}
{(j-i)\beta}
\,
\frac{1+(j-i+1)\beta}
{1+(j-i)\beta}
\cdots
\frac{B_i-B_j -1+(j-i+1)\beta}
{B_i-B_j -1+(j-i)\beta}
\\
& =
\frac{\prod_{j=m_t+1}^{m}
[(j-i+1)\beta]
\,
[(1+(j-i+1)\beta)]
\cdots
[(B_i-B_j -1+(j-i+1)\beta)]
}
{\prod_{k=m_t}^{m-1}
[(k-i+1)\beta]
\,
[(1+(k-i+1)\beta)]
\cdots
[(B_i-B_{k+1} -1+(k-i+1)\beta)]}
\\
& =
\frac{
[(m-i+1)\beta]
\,
[(1+(m-i+1)\beta)]
\cdots
[(B_i-B_m -1+(m-i+1)\beta)]
}
{
[(m_t-i+1)\beta]
\,
[(1+(m_t-i+1)\beta)]
\cdots
[(B_i-B_{m_t+1} -1+(m_t-i+1)\beta)]}
\times \\
& \times
\prod_{j=m_t+1}^{m-1}
\frac{
[(j-i+1)\beta]
\,
[(1+(j-i+1)\beta)]
\cdots
[(B_i-B_j -1+(j-i+1)\beta)]
}
{[(j-i+1)\beta]
\,
[(1+(j-i+1)\beta)]
\cdots
[(B_i-B_{j+1} -1+(j-i+1)\beta)]
}\;.
\end{split}
\end{equation}
Name the term in the last line to be H, we see  $H = 1$ unless $B_j\neq B_{j+1}$,
(i.e.,primary rows $j = m_k$). And notice that $B_{m_k +1}=B_{m_{k+1}}$,
we can count only over the primary rows.

As a result, we find
\begin{equation}
\begin{split}
&H = \\
&
\prod_{k=t+1}^{n-1}
\frac{
1
}
{
[(B_i-B_{m_{k+1}} -1+(m_k-i+1)\beta)]}
\,
\frac{
1
}
{[(B_i-B_{m_{k+1}} +(m_k-i+1)\beta)]}
\cdots
\frac{
1
}
{[(B_i-B_{m_k} +(m_k-i+1)\beta)]
}
\\
& =
\prod_{k=t+1}^{n-1}
\;
\prod_{j=B_{m_{k+1}}+1}^{B_{m_k}}
\frac{1}
{
 \Big[ (B_i - j)+ \beta (m_k - i +1) \Big]
}\;.
\end{split}
\end{equation}
Combine the above three equations, we obtain
\begin{equation}
\begin{split}
& R_1 \times L = \\
&
\frac{
[(m-i+1)\beta]
\,
[(1+(m-i+1)\beta)]
\cdots
[(B_i-B_m -1+(m-i+1)\beta)]
}
{
[(m_t-i+1)\beta]
\,
[(1+(m_t-i+1)\beta)]
\cdots
[(B_i-B_{m_t+1} -1+(m_t-i+1)\beta)]}
\times \\
& \times
\prod_{j=1}^{B_{m}}
 \Big[ (B_i - j)+ \beta (m - i +1) \Big]
\times
\prod_{j=B_{m_{t+1}}+1}^{B_{m_t}}
 \Big[ (B_i - j)+ \beta (m_t - i +1) \Big]
 \\
&=\prod_{j=1}^{B_i} [(B_i - j)+ \beta (m - i +1)]
=
\prod_{j=1}^{B_i} [( m\beta - \beta ( i - 1) +  j - 1 )]\;.
\end{split}
\end{equation}
This is equivalent to (\ref{lemma1a}), thus complete the proof of lemma 1.
\subsection{Proof of Lemma 2}
\paragraph{Lemma 2}
\begin{equation}
\prod_{i=1}^{N} (x - i \beta)_{B_i}
=\Big[ x - \beta \Big]_B
\end{equation}

\paragraph{Proof:}
Use (\ref{poch}),
we find
\begin{equation}
\begin{split}
&\prod_{i=1}^{N} (x - i \beta)_{B_i} \\
&=
\prod_{i=1}^{N}
\dfrac{\Gamma(x - i \beta+B_i)}{\Gamma(x - i \beta)}
=
\prod_{i=1}^{m}
\dfrac{\Gamma(x - i \beta+B_i)}{\Gamma(x - i \beta)}
=\prod_{i=1}^{m} (x - i \beta)(x - i \beta+1)\ldots(x - i \beta+B_i-1)=
\\
&=\prod_{i=1}^{m} \prod_{j=1}^{B_i} (x - i \beta +j-1)
=
\prod\limits_{(i,j) \in B} ( x - \beta- \beta ( i - 1) +  j - 1 )
=\Big[ x - \beta \Big]_B \;,
\end{split}
\end{equation}
where $m$ is the length of $B$.
\subsection{Proof of Lemma 3}
\paragraph{Lemma 3}
\begin{equation}
[ x ]_B
=
(-1)^{|B|} G_{B,\emptyset}(- x + 1 -\beta)
\end{equation}

\paragraph{Proof:}
\begin{equation}
[ x ]_B
=
\prod_{j=1}^{B_1} \prod_{i=1}^{ B'_j} ( x - \beta ( i - 1) +  j - 1 )
=
\prod_{j=1}^{B_1} \prod_{i=1}^{ B'_j} ( x - \beta ( B'_j - i) +  j - 1 )
=
(-1)^{|B|}G_{B,\emptyset}(- x + 1 -\beta)\;.
\end{equation}
The second equivalence is based on the fact that when $j$ is fixed,
 both $i - 1$ and $B'_j - i$ count from $0$ to $B'_j - 1$. \\ \\
\subsection{Proof of Lemma 4}
\paragraph{Lemma 4}
\begin{equation}
\begin{split}
\prod_{i=1}^{N_{1}}\prod_{j=1}^{N_{2}} \frac{\big(x+ 1 -(i +j)\beta \big)_{\beta}}
{\big(x+ 1 + A'_i + B_j -(i +j)\beta \big)_{\beta}}
=
\frac{(-1)^{\lvert B \rvert }[x - N_2\beta +1-\beta]_{A'} [x - N_1\beta+1-\beta]_B}
{G_{A,B}(x) G_{B,A}(-x )}
\end{split}
\end{equation}

Actually this lemma holds only for $\beta = 1$.  For this value, the equation
becomes,
\begin{equation}
\label{lemma4}
\prod_{i=1}^{N_{2}}\prod_{j=1}^{N_{1}} \frac{x+ 1 -( i +j) }
{x+1 + A'_j + B_i -(i +j)}
=
\prod_{(i,j)\in A}  \frac{x- N_2 +i -j }
{x + A'_j + B_i -i -j+1}
\prod_{(i,j)\in B}  \frac{x- N_1 -i +j }
{x - B'_j - A_i +i +j-1}\,.
\end{equation}
We have switched the name of $i$ and $j$
on the left hand side.

\paragraph{Proof: }
\underline{\bf  Step 1:} Proof for $B = \emptyset$. \\
The left hand side of (\ref{lemma4}) is,
\begin{equation}
\begin{split}
L_0&=\prod_{i=1}^{N_{2}}\prod_{j=1}^{N_{1}} \frac{x +1 -( i +j) }
{x +1 + A'_j -(i +j)}
=
\prod_{i=1}^{N_2}
\prod_{j=1}^{h}
\frac{x+1 -( i +j) }
{x +1 + A'_j -(i +j)}=
\\
&=
\prod_{i=1}^{N_2}
\prod_{j=1}^{h}
\prod_{k=1}^{A'_j}
\frac{x+ k - i -j }
{x +k+1 - i -j}
=
\prod_{j=1}^{h}
\prod_{k=1}^{A'_j}
\frac{x- N_2 +k  -j }
{x+k-j}
=
\prod_{(i,j)\in A}
\frac{x- N_2+i  -j }
{x+i-j}\;,
\end{split}
\end{equation}
where $h$ is the hight of $A$.

On the other hand, the right hand side of (\ref{lemma4}) becomes,
\begin{equation}
R_0=\prod_{(i,j)\in A}  \frac{x-N_2 +i -j }
{x  + (A'_j -i) -j+1}
=
\prod_{(i,j)\in A}
\frac{x-N_2+i  -j }
{x+i-j}\;.
\end{equation}
We see $L_0=R_0$, the equation (\ref{lemma4}) holds with $B = \emptyset$.

\begin{figure}[htbp]
\begin{center}
\includegraphics{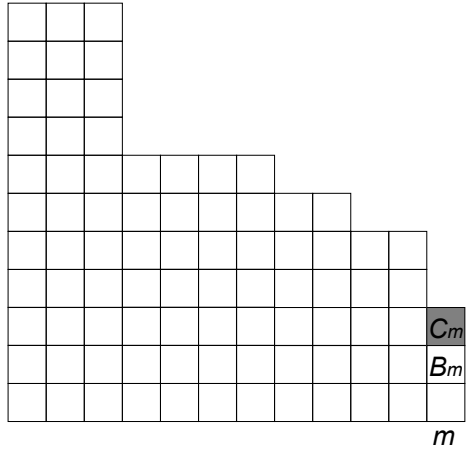}
\end{center}
\caption{Construction of $C$. The white cells stands for
 $B$ , while $C$ has one extra cell (marked in black) than
$B$ in the last column.}
\label{fig:3}
\end{figure}

\underline{\bf Step 2:} Induction for other cases. Suppose (\ref{lemma4}) is valid for $B$. As shown in Figure \ref{fig:3},
let us construct $C$ which has only one cell difference from $B$: $C_m=B_m +1$, $B'_{B_m +1}=m -1$, $C'_{B_m +1}=m $, with $m$ the length of $B$.
(Notice that the special case $B_m=0$ means $C_m$ starts from a new column,
thus we can build any diagram from zero).
\\
so we just need to prove that
\begin{equation}
\begin{split}
\label{lemma4a}
&\prod_{i=1}^{N_{2}}\prod_{j=1}^{N_{1}} \frac{x+1 -( i +j) }
{x +1 + A'_j + C_i -(i +j)}
\\
&=
\prod_{(i,j)\in A}  \frac{x-N_2 +i -j }
{x  + A'_j + C_i -i -j+1}
\prod_{(i,j)\in C}  \frac{x-N_1 -i +j }
{x  - C'_j - A_i +i +j-1}\;.
\end{split}
\end{equation}
The left hand side of (\ref{lemma4a}) is
\begin{equation}
\begin{split}
L&=\prod_{i=1}^{N_{2}}\prod_{j=1}^{N_{1}} \frac{x+1 -( i +j) }
{x +1 + A'_j + C_i -(i +j)}
\\
&=
\prod_{i=1}^{N_{2}}\prod_{j=1}^{N_{1}}  \frac{x+1 -( i +j) }
{x +1 + A'_j + B_i -(i +j)}
\prod_{j=1}^{N_{1}} \frac{x +1 + A'_j + B_m -(m +j) }
{x +1 + A'_j + B_m +1 -(m +j)}\;.
\end{split}
\end{equation}
The first term on the right hand side of (\ref{lemma4a}) is
\begin{equation}
\begin{split}
R_1&=
\prod_{(i,j)\in A}  \frac{x - N_2 +i -j }
{x  + A'_j + C_i -i -j+1}
\\
&=
\prod_{(i,j)\in A}  \frac{x- N_2 +i -j }
{x  + A'_j + B_i -i -j+1}
\prod_{j=1}^{A_m} \frac{x +1 + A'_j + B_m -(m +j) }
{x +1 + A'_j + B_m +1 -(m +j)}\,.
\end{split}
\end{equation}
And the second term becomes
\begin{equation}
\begin{split}
R_2&=
\prod_{(i,j)\in C}  \frac{x-N_1 -i +j }
{x  - C'_j - A_i +i +j-1}
\\
&=
\prod_{(i,j)\in B}  \frac{x-N_1 -i +j }
{x  - C'_j - A_i +i +j-1}
\times
\frac{x-N_1 -m +B_m +1 }
{x  - A_m +B_m}
\\
&=
\frac{x-N_1 -m +B_m +1 }
{x  - A_m +B_m}
\times
\prod_{(i,j)\in B}  \frac{x-N_1 -i +j }
{x  - B'_j - A_i +i +j-1}
\prod_{i=1}^{m-1}  \frac{x  - m - A_i +i + B_m +1 }
{x  - m - A_i +i + B_m}\;.
\end{split}
\end{equation}
Since we have assumed the equation (\ref{lemma4})is correct for $B$,
we only need to proof
\begin{equation}
\begin{split}
&\prod_{j=1}^{N_1} \frac{x +1 + A'_j + B_m -(m +j) }
{x +1 + A'_j + B_m +1 -(m +j)}
\\
&=
\prod_{j=1}^{A_m} \frac{x +1 + A'_j + B_m -(m +j) }
{x +1 + A'_j + B_m +1 -(m +j)}
\\
&\times
\frac{x-N_1 -m +B_m +1 }
{x  - A_m +B_m}
\times
\prod_{i=1}^{m-1}  \frac{x  - m - A_i +i + B_m +1 }
{x  - m - A_i +i + B_m}\;,
\end{split}
\end{equation}
which is equivalent to
\begin{equation}
\begin{split}
\label{lemma4b}
&\prod_{j=A_m +1}^{N_1} \frac{x  + A'_j -j + B_m -m +1 }
{x  + A'_j -j + B_m -m +2 }
=
\frac{x- N_1 -m +B_m +1 }
{x  - A_m +B_m}
\times
\prod_{i=1}^{m-1}  \frac{x  - m - A_i +i + B_m +1 }
{x  - m - A_i +i + B_m}\,.
\end{split}
\end{equation}
The left hand side of the above transforms to
\begin{equation}
\begin{split}
\label{lemma4c}
L'&=
\prod_{j=A_m +1}^{N_1} \frac{x  + A'_j -j + B_m -m +1 }
{x  + A'_j -j + B_m -m +2 }
\\
&=
\prod_{j=h +1}^{N_1} \frac{x   -j + B_m -m +1 }
{x  -j + B_m -m +2 }
\prod_{j=A_m +1}^{h} \frac{x  + A'_j -j + B_m -m +1 }
{x  + A'_j -j + B_m -m +2 }
\\
&=
\frac{x-N_1   + B_m -m +1 }
{x   + B_m -m +1-h }
\prod_{j=A_m +1}^{h} \frac{x  + A'_j -j + B_m -m +1 }
{x  + A'_j -j + B_m -m +2 }\;.
\end{split}
\end{equation}
Here $h$ is again the hight of $A$. Name the second term of the last line as $L'_1$,
\begin{equation}
\begin{split}
\label{lemma4d}
L'_1&=
\prod_{j=A_m +1}^{h} \frac{x  + A'_j -j + B_m -m +1 }
{x  + A'_j -j + B_m -m +2 }
\\
&=
\prod_{j=A_m +1}^{h} \frac{x  + A'_j -j + B_m -m +1 }
{x  -j + B_m -m +1 }
\frac{x  -j + B_m -m +1 }
{x  + A'_j -j + B_m -m +2 }
\\
&=
\prod_{j=A_m +1}^{h}
\bigg(
\prod_{i=1}^{A'_j}
\frac{x  -j + B_m -m+i +1 }
{x  -j + B_m -m+i }
\prod_{i=0}^{A'_j}
\frac{x  -j + B_m -m+i +1 }
{x  -j + B_m -m+i+2 }
\bigg)
\\
&=
\prod_{j=A_m +1}^{h}
\frac{x  -j + B_m -m+1  }
{x  -j + B_m -m+2 }
\times
\\
&\times
\prod_{j=A_m +1}^{h}
\prod_{i=1}^{A'_j}
\bigg(
\frac{x  -j + B_m -m+i +1 }
{x  -j + B_m -m+i }
\frac{x  -j + B_m -m+i +1 }
{x  -j + B_m -m+i+2 }
\bigg)\;,
\end{split}
\end{equation}
This time we call the last term of the last line as $L_3$.

The second term of the right hand side of (\ref{lemma4b}) has the form
\begin{equation}
\begin{split}
R'_2&=
\prod_{i=1}^{m-1}  \frac{x  - m - A_i +i + B_m +1 }
{x  - m - A_i +i + B_m}
\\
&=
\prod_{i=1}^{m-1}  \frac{x  - m - A_i +i + B_m +1 }
{x  - m  +i + B_m +1}
\frac{x  - m  +i + B_m +1}
{x  - m - A_i +i + B_m  }
\\
&=
\prod_{i=1}^{m-1}
\bigg(
\prod_{j=1}^{A_i}
 \frac{x  - m +i + B_m-j +1 }
{x  - m +i + B_m-j +2 }
\prod_{j=0}^{A_i}
 \frac{x  - m +i + B_m-j +1 }
{x  - m +i + B_m-j  }
\bigg)
\\
&=
\prod_{i=1}^{m-1}
 \frac{x  - m +i + B_m +1 }
{x  - m +i + B_m  }
\times
\\
&\times
\prod_{i=1}^{m-1}
\prod_{j=1}^{A_i}
\bigg(
 \frac{x  - m +i + B_m-j +1 }
{x  - m +i + B_m-j +2 }
 \frac{x  - m +i + B_m-j +1 }
{x  - m +i + B_m-j  }
\bigg)\,,
\end{split}
\end{equation}
\begin{figure}[htbp]
\centering
\includegraphics{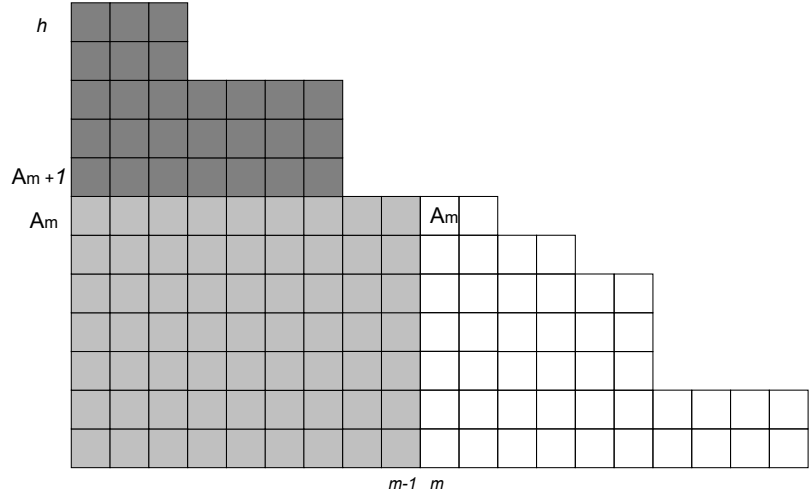}
\caption{
$\prod_{i=1}^{m-1}\prod_{j=1}^{A_i}$ is represented by the area marked by grey and black, while
$\prod_{j=A_m +1}^{h}\prod_{i=1}^{A'_j}$ is represented only by the black cells. Their difference,
the grey cells, stands for $\prod_{i=1}^{m-1}\prod_{j=1}^{A_m}$, which leads to the following
equation.
}
\label{fig:4}
\end{figure}\\
so we find (see Figure \ref{fig:4})
\begin{equation}
\begin{split}
\label{lemma4e}
\frac{R'_2 }
{L_3}&=
\prod_{i=1}^{m-1}
 \frac{x  - m +i + B_m +1 }
{x  - m +i + B_m  }
\times
\\
&\times
\prod_{i=1}^{m-1}
\prod_{j=1}^{A_m}
\bigg(
 \frac{x  - m +i + B_m-j +1 }
{x  - m +i + B_m-j +2 }
 \frac{x  - m +i + B_m-j +1 }
{x  - m +i + B_m-j  }
\bigg)
\\
&=
\prod_{i=1}^{m-1}  \frac{x  - m - A_m +i + B_m +1 }
{x  - m - A_m +i + B_m}\;.
\end{split}
\end{equation}
Combine (\ref{lemma4c}), (\ref{lemma4d}) and (\ref{lemma4e}), it is straightforward to find that
 (\ref{lemma4b}) is tenable, thus complete the proof. \\ \\

\subsection{Proof of Lemma 5}
\paragraph{Lemma 5}
\begin{equation}
\begin{split}
\label{trick5}
&\prod_{i=1}^{N_{1}}\prod_{j=1}^{N_{2}} \frac{\big(x+ 1 -(i +j)\beta \big)_{\beta}}
{\big(x+ 1 + B_j -(i +j)\beta \big)_{\beta}}
=
\frac{[x - N_1\beta +1 -\beta]_B}
{[x +1 -\beta]_B} \;,
\\
&\prod_{i=1}^{N_{1}}\prod_{j=1}^{N_{2}} \frac{\big(x+ 1 -(i +j)\beta \big)_{\beta}}
{\big(x+ 1 + A'_i -(i +j)\beta \big)_{\beta}}
=
\frac{[x - N_2\beta +1 -\beta]_{A'}}
{[x +1 -\beta]_{A'}}
\end{split}
\end{equation}

These are actually the special case of Lemma 4, but hold for arbitrary $\beta$.

\paragraph{Proof:}
For the first statement, we have
\begin{equation}
\begin{split}
&L
=
\prod_{i=1}^{N_{1}}\prod_{j=1}^{N_{2}} \frac{\big(x+ 1 -(i +j)\beta \big)_{\beta}}
{\big(x+ 1 + B_j -(i +j)\beta \big)_{\beta}}
\\
&=
\prod_{i=1}^{N_1}
\prod_{j=1}^{m}
\frac{\big(x +1 -(i +j)\beta \big)\big(x +2 -(i +j)\beta \big) \dots
\big(x-(i +j-1)\beta \big)}
{\big(x+1+ B_j -(i +j)\beta \big)\big(x+ 2+ B_j -(i +j)\beta \big) \dots
\big(x+ B_j -(i +j-1)\beta \big)}
\\
&=
\prod_{i=1}^{N_1}
\prod_{j=1}^{m}
\prod_{k=1}^{B_j}
\frac{x+k -(i +j)\b}
{x +k -(i +j-1)\b}
=
\prod_{j=1}^{m}
\prod_{k=1}^{B_j}
\frac{x-N_1\b+k-j\b }
{x+k -j\b }=
\\
&=
\prod_{(i,j)\in B}
\frac{x-N_1\b- i\b +j }
{x- i\b+j }
=
\frac{[x -N_1\beta +1 -\beta]_B}
{[x + 1 -\beta]_B}
= R\;,
\end{split}
\end{equation}
where $m$ is the length of $B$.

The second statement can be proved in totally the same way.

\end{document}